\documentclass[fleqn,usenatbib]{mnras}
\usepackage{newtxtext,newtxmath}
\usepackage[T1]{fontenc}
\usepackage{ae,aecompl}
\usepackage{graphicx}	
\usepackage{amsmath}

\usepackage{amssymb}
\usepackage{times}
\usepackage{verbatim}
\usepackage{subfigure}
\usepackage{float}

\renewcommand\labelenumi{(\roman{enumi})}
\renewcommand\theenumi\labelenumi

\title[CGM and SF: synced breathing, entangled fate]{From large-scale environment to CGM angular momentum to star forming activities -- I: star-forming galaxies}
\author[S. Wang et al.]
{Sen Wang$^{1}$\thanks{E-mail: \url{wangsen19@mails.tsinghua.edu.cn}},
Dandan Xu$^{1}$\thanks{E-mail: \url{dandanxu@tsinghua.edu.cn}},
Shengdong Lu$^{1}$,
Zheng Cai$^{1}$,
Maosheng Xiang$^{2}$,
\and
Shude Mao$^{1,3}$,
Volker Springel$^{4}$,
Lars Hernquist$^{5}$
\\
\\
$^{1}$Department of Astronomy, Tsinghua University, Beijing 100084, China\\
$^{2}$Max-Planck-Institute for Astronomy, Koenigstuhl 17, 69117, Heidelberg, Germany\\
$^{3}$National Astronomical Observatories, Chinese Academy of Sciences, 20A Datun Road, Chaoyang District, Beijing 100101, China\\
$^{4}$Max-Planck-Institut f\"ur Astrophysik, Karl-Schwarzschild-Str. 1, D-85748, Garching, Germany\\
$^{5}$Harvard-Smithsonian Center for Astrophysics, 60 Garden Street, Cambridge, MA 02138, USA\\
}

\date{Accepted ***. Received ***; in original form ***}

\pubyear{2021}

\begin{document}
\label{firstpage}
\pagerange{\pageref{firstpage}--\pageref{lastpage}}
\maketitle
\begin{abstract} 
The connection between halo gas acquisition through the circumgalactic medium (CGM) and galaxy star formation has long been studied. In this series of two papers, we put this interplay within the context of the galaxy environment on large scales (several hundreds of kpc), which, to a certain degree, maps out various paths for galaxy interactions. We use the IllustrisTNG-100 simulation to demonstrate that the large-scale environment modulates the circumgalactic gas angular momentum, resulting in either enhanced (Paper I) or suppressed (Paper II) star formation inside a galaxy. In this paper (Paper I), we show that the large-scale environment around a star-forming galaxy is often responsible for triggering new episodes of star formation. Such an episodic star formation pattern is well synced with a pulsating motion of the circumgalactic gas, which, on the one hand receives angular momentum modulations from the large-scale environment, yielding in-spiralling gas to fuel the star-forming reservoir, while, on the other hand, is affected by the feedback activities from the galaxy centre. As a result, a present-day star-forming galaxy may have gone through several cycles of star-forming and quiescent phases during its evolutionary history, with the circumgalactic gas carrying out a synchronized cadence of ``breathing in and out'' motions out to $\sim 100$ kpc.
\end{abstract}


\section{Introduction}
\label{sec:introduction}
The evolution of galaxies is tightly linked to the gas reservoirs surrounding them, known as the circumgalactic medium (CGM). The CGM is a massive multiphase component residing inside galaxy halos. The acquisition of this reservoir is achieved through gas accretion from the intergalactic medium (IGM) along filaments of the cosmic web. Star-forming galaxies need continuous replenishment of fresh (low-metallicity) gas within the CGM accreted from the IGM \citep{Fox2017}. Evidence comes from the presence of low-metallicity stars (metal-poor G-dwarf population) \citep{van_den_Bergh_1962,Schmidt_1963,Larson_1972a,Larson_1972b,Lynden-Bell_1975,Pagel&Patchett_1975,Sommer-Larsen_1991,Casuso&Beckman_2004,Woolf&West_2012}, observations of high-column density low-metallicity IGM absorbers (e.g., \citealt{Lehner_et_al_2013,Hafen_et_al_2017}), as well as sustainable star formation (SF) in galaxies \citep{Daddi_et_al_2010,Freundlich_et_al_2013,Genzel_et_al_2015,Tacconi_et_al_2010,Tacconi_et_al_2013,Tacconi_et_al_2018,Scoville_et_al_2016,Scoville_et_al_2017,Schinnerer_et_al_2016,Saintonge_et_al_2013,Saintonge_et_al_2016,Saintonge_et_al_2017}. 

Gas accretion from the IGM via the CGM not only contributes to metal-poor pristine gas to fuel star formation (e.g., \citealt{Dekel_et_al_2009,van_de_Voort_et_al_2011}) but also brings angular momentum to the galaxy (e.g., \citealt{Stewart_et_al_2011b,Stewart_et_al_2013,Stewart_et_al_2017,Danovich_et_al_2015}). In particular, more recent theories emphasized the importance of the filamentary cold-stream/cold-flow accretion. In this process, high angular-momentum gas originates from the cosmic web and flows onto the (outskirts of) galaxies in an in-spiraling fashion, in contrast to the conventional isotropic hot-mode accretion mode, where gas has been shock-heated to the virial temperature before cooling onto the galaxy. Angular momenta in the cold-accretion case are much higher than in the shock-heated case (e.g., \citealt{Keres_et_al_2005,Keres_et_al_2009,Dekel&Birnboim_2006,Dekel_et_al_2009,Brooks_et_al_2009,Stewart_et_al_2011a,Stewart_et_al_2011b,Stewart_et_al_2017,van_de_Voort_et_al_2011,Danovich_et_al_2015,Tumlinson_et_al_2017}).

Rich and complex kinematic structures of circumgalactic gas around galaxies have already been seen in cosmological simulations  \citep{Nelson_et_al_2013,Ford_et_al_2014,Angles-Alcazar_et_al_2017, Oppenheimerr_2018,Hafen_et_al_2019,Nelson_et_al_2020}, such as
extended, warped and thick discs due to cosmological gas accretion \citep{Stewart_et_al_2011b, Stewart_et_al_2013}, gas associated with accreting satellites \citep{Shao_et_al_2018} and stream accretion from the cosmic web \citep{Dekel_et_al_2009,Danovich_et_al_2015}. In particular, high-angular-momentum co-rotating (with respect to stellar discs) circumgalactic gas associated with simulated galaxies is found to extend to $\sim$ 100 kpc  \citep{El-Badry_et_al_2018,Ho_et_al_2020, Huscher_et_al_2021,DeFelippis_et_al_2021}.

Observationally, high-angular-momentum CGM coherently rotating around galaxy discs has also been seen for decades. In our Milky Way, this process is evidenced by detections of the highest velocity clouds \citep{Zheng_et_al_2015} and rotating hot halo gas ($180\,\mathrm{km\,s^{-1}}$) from X-ray measurements of OVII \citep{Hodges-Kluck_et_al_2016}. Those in the local Universe include detections of extended HI discs out to several tens to hundreds of kpc, giant low-surface brightness galaxies and low-metallicity high-angular momentum gas \citep{Bosma_1981,Bothun_et_al_1987,Matthews_et_al_2001,Putman_et_al_2009,Bigiel_et_al_2010,Spavone_et_al_2010,Kreckel_et_al_2011,Holwerda_et_al_2012,Hagen_et_al_2016,Wang_et_al_2016}. At lower redshifts, thick and extended co-rotating gaseous discs (e.g., \citealt{Steidel_et_al_2002, Diamond-Stanic_et_al_2016}) and in-spiraling gas towards inner discs (e.g., \citealt{Ho_et_al_2017}) have also been detected using absorption lines in spectra of background quasars. 

Absorption line studies of galaxies' CGM at even higher redshifts ($z\sim 0.5-1.5$) have unveiled interesting gas kinematics: gas inflow tends to happen along a galaxy’s major axis and co-rotating with the galactic disc; while gas outflow tends to happen more along the minor axis \citep{Kacprzak_et_al_2010,Kacprzak_et_al_2012,Bouche_et_al_2012,Bouche_et_al_2013,Bouche_et_al_2016,Crighton_et_al_2013,Nielsen_et_al_2015,Danovich_et_al_2015,Diamond-Stanic_et_al_2016,Bowen_et_al_2016,Ho_et_al_2017,Martin_et_al_2019,Zabl_et_al_2019}. At even higher redshifts ($z \sim 2-3$), high-angular-momentum gas on large scales flowing in along cosmic filaments has been detected, and in some cases seen to kinematically link to extended proto-galactic discs \citep{Martin_et_al_2014, Martin_et_al_2015, Martin_et_al_2016, Prescott_et_al_2015, Battaia18LyAEmitAtZ3, Cai19LyAEmitAtZ2}. Non-zero angular momentum has also been identified in metal-enriched CGM gas (\citealt{Cai17LyANebAtZ2}). All these suggest strong connections between the angular-momentum-carrying circumgalactic gas at larger radii and star-forming discs on smaller scales.  

In this series of two papers, we examine the interplay between the circumgalactic gas motion and galaxy star formation within the context of the large-scale environment (of hundreds of kpc), which, to a certain degree, maps out various paths for galaxy interactions including mergers and fly-bys. It has been shown in both observations and theory that the environment closely links to star formation, for example, galaxy interaction and major mergers can significantly enhance central star formation or even trigger starbursts inside galaxies (e.g., \citealt{Barnes_et_al.(1992),Barnes_et_al.(1991),Barnes_et_al.(1996),Mihos_et_al.(1996), Barnes_et_al.(2002),Naab_et_al.(2003),Bournaud_et_al.(2005),Bournaud_et_al.(2007),Johansson_et_al.(2009a),Tacchella_et_al.(2016),Rodriguez-Gomez_et_al.(2017)}). In this series,  we use the IllustrisTNG-100 simulation \citep{Marinacci_et_al.(2018),Naiman_et_al.(2018),Springel_et_al.(2018),Nelson_et_al.(2018),Nelson_et_al.(2019b),Pillepich_et_al.(2018b),Pillepich_et_al.(2019)} to show specifically that the large-scale environment can modulate the CGM angular momentum, resulting in either episodic (this work, Paper I) or suppressed (Paper II, \citealt{Lu21PaperIIQuench}) star formation inside galaxies.

In this work, specifically, we demonstrate that galaxy interactions (through the CGM) play an essential role in triggering new episodes of star formation in present-day star-forming galaxies. In this process, the circumgalactic gas receives angular-momentum modulations from the large-scale environment, in-spirals onto the galaxy and supplies the central cold gas reservoir, starting new episodes of star formation. On the other hand, the active galactic nuclei (AGN) and stellar feedback activities inside a galaxy also affect the circumgalactic gas through injecting metals as well as energy and momentum into the CGM. This not only enriches and heats up the circumgalactic gas at larger radii, but also prevents the cold gas from efficiently flowing back to the galaxy, until the feedback strength diminishes as the star-forming activity subsides. A new cycle of star formation then starts, when the cold circumgalactic gas once again cools down to the galaxy. As a consequence, a present-day star-forming galaxy can have gone through several rounds of star formation episodes with quiescent phases in between, in a rhythm synced with the motion of the circumgalactic gas at larger radii. In particular, the cold ($\sim 10^4$ K) and hot ($>10^5$ K) components of the circumgalactic gas carry out different kinematics. Together they produce a ``breathing in and out'' motion out to $\sim 100$ kpc, with a modulation shared by episodic star-forming and quenching cycles inside the galaxy.

These synced pulsations and entangled fates between the circumgalactic gas and central star formation, as modulated by both the external environment and the internal feedback, result in a clear observational consequence: an episodic star formation history. We predict that some of these past star-formation episodes with clear quiescent phases in between can be extracted from spatially resolved stellar populations for the Milky Way and/or nearby galaxies.

The paper is organized as follows. In Section~\ref{sec:method}, we summarize details of the simulation, galaxy selection, and CGM property calculations in this study. 
In Section~\ref{sec:breathing}, we present an episodic ``rise and fall'' pattern of star formation along cosmic time, in sync with a ``breathing in and out'' motion of the circumgalactic gas. In particular, motions of the cold and the hot circumgalactic gas are presented in Section~\ref{sec:cold} and \ref{sec:hot}, respectively. With an understanding of the cold and hot CGM motion, we then in  Section~\ref{sec:MergingEnvironment}, present three pieces of key evidence that galaxy interactions are often responsible for new epochs of star formation. In Section~\ref{sec:observation}, observational predictions are proposed. Conclusions and some further discussion are finally given in Section~\ref{sec:conclusion}. In this work, we adopt the same cosmology as those used in the IllustrisTNG simulation (based on results of the Planck experiment, \citealt{Planck_Collaboration(2016)}). Specifically, a flat universe geometry is assumed, with a total matter density of $\Omega_{\rm m} = 0.3089$, a cosmological constant of $\Omega_{\Lambda} = 0.6911$, a baryonic density of $\Omega_{\rm b} = 0.0486$, and a Hubble constant $h = H_0/(100\,{\rm km s}^{-1} {\rm Mpc^{-1}}) = 0.6774$.

\section{Methodology}
\label{sec:method}
\subsection{The simulation}

\textit{The Next Generation Illustris Simulations} (IllustrisTNG, TNG
hereafter; \citealt{Marinacci_et_al.(2018),Naiman_et_al.(2018),Springel_et_al.(2018),Nelson_et_al.(2018),Nelson_et_al.(2019b),Pillepich_et_al.(2018b),Pillepich_et_al.(2019)}) are a set of magneto-hydrodynamic cosmological simulations using the moving-mesh code \textsc{arepo} \citep{Springel(2010)} for galaxy formation and evolution. In this study, we use IllustrisTNG-100 (TNG-100 hereafter), which was run within a cubic box of $110.7\,\mathrm{Mpc}$ side length and with a mass resolution of $1.4\times10^6\,{\rm M_{\odot}}$ and $7.5\times10^6\,{\rm M_{\odot}}$ for the baryonic and dark matter, respectively, and a gravitational softening length of $0.5\,\mathrm{h^{-1}kpc}$ for the dark matter and stellar particles. The {\sc subfind} algorithm \citep{Springel_et_al.(2001),Dolag_et_al.(2009)} was used to identify galaxies in their host halos. General galaxy properties were calculated and publicly released by the TNG collaboration\footnote{\url{http://www.tng-project.org/data/}}\citep{Nelson_et_al_2019a}.

\subsection{Galaxy sample selection}
\label{sec:method1}
The galaxy samples used in this work are constructed based on the samples from \citet{Lu21HotDisk, Lu2021ColdElliptical}, in which classical late- and early-type galaxies were identified according to their specific star-formation rates (SFR), bulge-to-total luminosity ratios and circular orbit fractions. We take their disc galaxy samples of two stellar mass ranges at $z=0$: a lower-mass sample with $9.5<\log\,M_{\ast}/\mathrm{M_{\odot}}<10.3$ (referred to as normal discs - NDs) and a more massive sample with $10.3<\log\,M_{\ast}/\mathrm{M_{\odot}}<11.2$ (referred to as normal massive discs - NMDs). Each sample contains about a hundred galaxies, together composing our present-day disc galaxy sample. In addition, we also visually inspected the redshift evolution and merger history of each galaxy, and finally selected 15 present-day disc galaxies (6 NDs $+$ 9 NMDs) that exhibit typical patterns of the pulsating CGM motion over the cosmic time. We use their progenitor properties since $z=1$ for statistical analysis.

\subsection{Defining the cold and hot circumgalactic gas}
\label{sec:method2}
In order to evaluate the CGM properties, we first follow the practice of \citet{DeFelippis_et_al_2020} and define the region outside twice the half-stellar-mass radius as the CGM domain; i.e., $r>2R_{\rm hsm}$. Key properties of the circumgalactic gas, including mass (fraction), metallicity, radial velocity, velocity anisotropy, and angular momentum, are evaluated at several distances from the galactic centre, in particular, within $5R_{\rm hsm}$ (roughly corresponding to $15-25\,\rm kpc$ at $z=0$), $10R_{\rm hsm}$ and fixed distances at 20, 50 and 100 kpc. Due to bi-modal gas accretion and the multi-phase nature of the CGM, we explicitly distinguish between the cold and the hot CGM phases to study their different motion and properties within the presented theoretical framework. Specifically, we divide the circumgalactic gas into a cold phase that has a temperature of $T = 10^4\,\mathrm{K}-2 \times 10^4\,\mathrm{K}$ and a hot phase that has a temperature of $T > 10^5\,\mathrm{K}$, which roughly correspond to the coldest 30 per cent and the hottest 30 per cent CGM at $T > 10^4\,\mathrm{K}$, respectively.

\section{Episodic star formation synced with breathing circumgalactic gas}
\label{sec:breathing}

\label{sec:red_evolv}

\begin{figure}
\centering
\includegraphics[width=1\columnwidth]{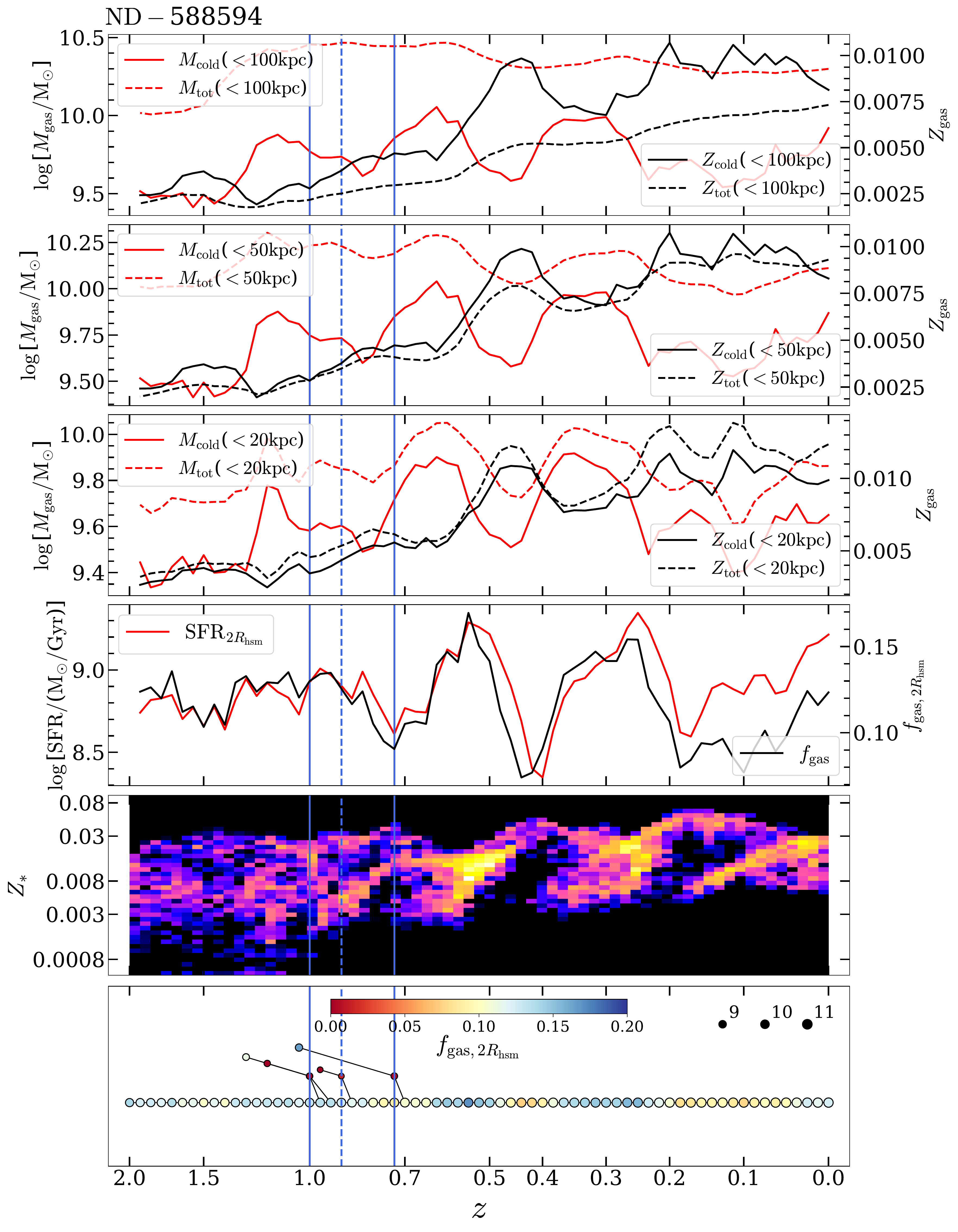} 
\caption{The redshift evolution of galaxy gas and star-formation properties since $z\sim 2$ for an example galaxy (ID-588594). The first three rows present the evolution of the gas mass (red) and the gas metallicity (black) for cold (solid) and total gas (dashed) within three apertures ($100\,\rm kpc$, $50\,\rm kpc$ and $20\,\rm kpc$ from the first to the third rows). The fourth row shows the redshift evolution of the central gas fraction ($f_{{\rm gas},2R_{\rm hsm}}$, black) within 2$R_{\rm hsm}$ and SFR (red). The fifth row shows the stellar intensity map in the redshift-star metallicity ($z - Z_{\ast}$) plane, of which the brighter part represents higher SFR at a given time and with a given metallicity. The last panel displays the merger history of the galaxy; each epoch (circle) is color-coded by the gas mass fraction within 2$R_{\rm hsm}$. The vertical lines mark the redshifts of recorded galaxy mergers. Red and blue colors correspond to gas-poorer and gas-richer mergers, respectively. The solid and the dashed lines indicate major or minor mergers, respectively (see \citealt{Lu2021ColdElliptical} for definitions).}
\label{fig:Evolution_breathing}
\end{figure}

In Fig.\,\ref{fig:Evolution_breathing}, we show the redshift evolution of one example galaxy, which is a present-day normal star-forming disc galaxy and has a subhalo ID\footnote{The simulation gives each galaxy a unique index at each snapshot, referred to as the ``subhalo ID''. The ``galaxy IDs'' throughout this paper refer to the subhalo IDs at $z=0$.} of 588594 at $z=0$. That of eight other traced galaxies can be found in Figs.\,\ref{fig:example_track_ND} and \ref{fig:example_track_NMD} in Appendix~\ref{apd:examples}. The first three panels present the redshift evolution of the mass and metallicity of the cold and total circumgalactic gas enclosed within 3D radii of 100 kpc, 50 kpc and 20 kpc. The fourth panel shows the redshift evolution of the central gas fraction ($f_{{\rm gas},2R_{\rm hsm}}$, black) and SFR (red). The fifth panel presents the intensity of newly formed stars at a given redshift $z$ and with a given stellar metallicity $Z_{\ast}$. The last panel shows the merger history of the galaxy. As can be seen, with time the galaxy's SFR rises and falls with a specific rhythm, which syncs with the evolution of the cold-gas mass and metallicity enclosed within a given aperture. The peaks of the SFR slightly trail behind {\it in time} the peaks of the enclosed gas mass, which roughly correspond to troughs of the gas metallicity. This holds from small scales to as far as $\sim 100$ kpc, and in particular so for the cold gas. 

\begin{figure}
\centering
\includegraphics[width=1\columnwidth]{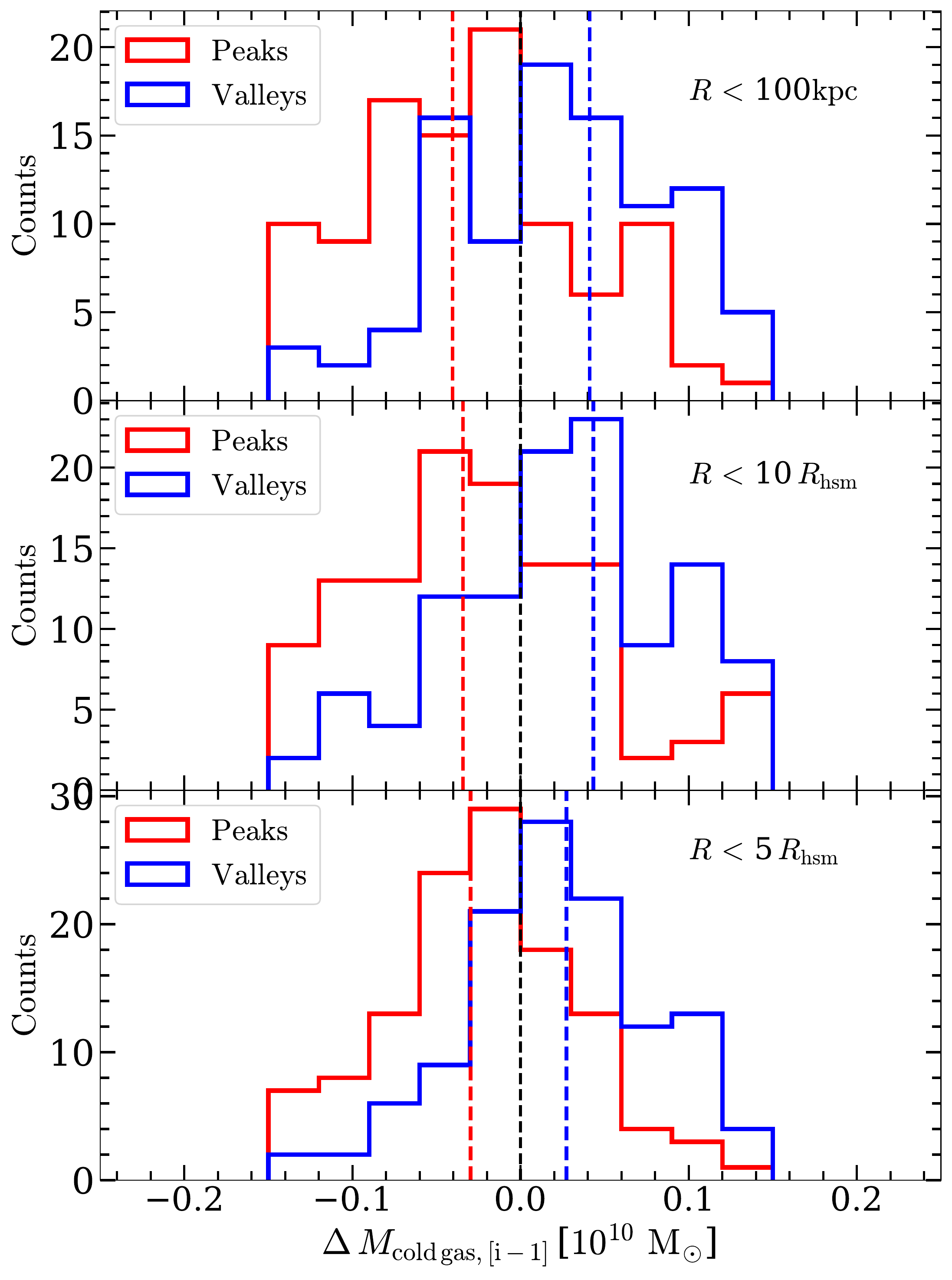}
\caption{Distributions of $\Delta M_{\rm cold\,gas}$ at the peaks (red) and valleys (blue) of SFR (as shown in Fig.\,\ref{fig:Evolution_breathing}) for the selected sample of 15 galaxies and their progenitors since $z=1$. Here $\Delta M_{\rm cold\,gas,\,[i-1]}\equiv M_{\rm cold\,gas,\,[i]}-M_{\rm cold\,gas,\,[i-1]}$ (see Section~\ref{sec:breathing} for details); peaks refer to the points satisfying $\Delta \rm SFR_{\rm i}\equiv \rm SFR_{\rm i+1}-SFR_{\rm i}<0$ and $\Delta \rm SFR_{\rm i-1}\equiv \rm SFR_{\rm i}-SFR_{\rm i-1}>0$, while valleys correspond to where $\Delta \rm SFR_{\rm i}\equiv \rm SFR_{\rm i+1}-SFR_{\rm i}>0$ and $\Delta \rm SFR_{\rm i-1}\equiv \rm SFR_{\rm i}-SFR_{\rm i-1}<0$. The three panels show the results at 100 kpc, 10 $R_{\rm hsm}$ ($\sim 30\,\rm kpc$ for the ND samples and $\sim 50\,\rm kpc$ for the NMD samples) and 5 $R_{\rm hsm}$ from top to bottom. In each panel, the red and blue dashed lines indicate the median values of $\Delta M_{\rm cold\,gas}$ for peak and valley samples, respectively. The black dashed line indicates $\Delta M_{\rm cold\,gas}=0$.}
\label{fig:relation between cold gas and sfr}
\end{figure}

A universal relation between the SFR and the enclosed gas mass, as revealed by the example galaxy in Fig.~\ref{fig:Evolution_breathing}, can be clearly seen in the traced samples of 15 galaxies as shown in Fig.\,\ref{fig:relation between cold gas and sfr}. Here we first define a SFR difference between two adjacent snapshots as $\Delta {\rm SFR}_{\rm i} \equiv {\rm SFR}_{\rm i+1} - {\rm SFR}_{\rm i}$, evaluated within $2\,R_{\rm hsm}$. We also define a CGM cold-gas mass difference within a given aperture as $\Delta M_{\rm cold\,gas,\,[i]}(\leqslant R) \equiv M_{\rm cold\,gas,\,[i+1]}(\leqslant R) -M_{\rm cold\,gas,\,[i]}(\leqslant R) $. In both cases, $\rm i$ represents the $\rm i^{\rm th}$ snapshot with an increasing time order; i.e., the $\rm (i-1)^{\rm th}$ snapshot is prior (in terms of cosmic time) to the $\rm i^{\rm th}$, further prior to the $\rm (i+1)^{\rm th}$ etc. The mass differences are calculated within three different 3D aperture radii, i.e., $5\,R_{\rm hsm}$, $10\,R_{\rm hsm}$ and 100 kpc. We then evaluate the relation at those snapshots where $\Delta {\rm SFR}_{\rm i} \times \Delta {\rm SFR}_{\rm i-1} < 0$, i.e., when the instant SFR reaches a local peak ($\Delta {\rm SFR}_{\rm i}<0$) or a local valley ($\Delta {\rm SFR}_{\rm i}>0$) at the $\rm i^{\rm th}$ snapshot. We note that we use $\Delta M_{\rm cold\,gas,\,[i-1]}$ at one snapshot prior to that of $\Delta {\rm SFR}_{\rm i}$ such that the relation between the two can indicate how changing the cold circumgalactic gas (contained within a fixed aperture) would influence the SFR in the following time step. From Fig.~\ref{fig:relation between cold gas and sfr}, it is clearly seen that an SFR trough (peak) often associates with a previous increase (decrease) of the cold gas content, which stops a further decline (surge) and leads to a rise (fall) of the SFR. More importantly, this behavior is true across a large radial range, and more so on larger scales, indicating the cold gas coming down from larger radii. 

As shown in Section~\ref{sec:MergingEnvironment}, distinct star-forming episodes (as seen in Fig.\,\ref{fig:Evolution_breathing}) are very often connected to merger or fly-by interactions. The resulted external gas supplements bring in low-metallicity gas, causing the new episodes to begin from low metallicity statuses, as is shown in the $z_{\rm SF} - Z_{\ast}$ phase diagram (in the fifth panel of Fig.\,\ref{fig:Evolution_breathing}). With time, the cold gas reservoir gradually builds up, star formation gets stronger and stronger, and so do the AGN and stellar feedback activities, which heat up the environment, pollute the ambient gas, and lift the hot and metal-enriched gas to large radii. As is shown in Section~\ref{sec:hot}, the outward moving hot gas that originates from the AGN feedback counter-balances the infalling cold gas at around $10-15\,R_{\rm hsm}$, allowing the cold circumgalactic gas to remain outside. As this process carries on, star formation and feedback activities continue metal-enriching the environment while consuming the gradually diminishing cold gas reservoir. The cold-gas mass and the metallicity therefore present opposite evolutionary trends in time.

It is worth noting that the newly incoming lower-metallicity gas is often first present at larger radii of the galaxy, while the gas giving rise to more rapid star-formation that happens in the central/bulge region of the galaxy is often highly metal-rich with a typical value of several solar metallicity $Z_{\odot}$. This effect shows up itself as a higher-metallicity branch in the $z_{\rm SF} - Z_{\ast}$ phase diagram, as can be seen in Figs.\,\ref{fig:example_track_ND} and \ref{fig:example_track_NMD} in the Appendix~\ref{apd:examples}. 
 
For our star-forming disc samples, we estimate the depletion time of HI (due to star formation) within a radius of $2\,R_{\rm hsm}$ from the galaxy center, which is typically between 0.5 to 2 Gyrs (with a median value around 1 Gyr), consistent with observational constraints from \citet{LinXue21_GasDepletionTime}. Here the depletion time is calculated as the HI mass divided by the SFR within $2\,R_{\rm hsm}$. With time, star formation and stellar/AGN feedback activities will eventually exhaust a significant amount of the existing cold gas reservoir, if there is no further gas supply. As the central gas fraction drops, the star formation and feedback activities would become weaker and weaker. The reduced feedback strength would then allow the (low-angular-momentum) cold circumgalactic gas (as often perturbed by galaxy interactions, see Section~\ref{sec:MergingEnvironment}, also see \citealt{Peng&Renzini}) to flow in and fuel further star formation -- a new episode thus begins. 

\begin{figure}
\centering
\includegraphics[width=1\columnwidth]{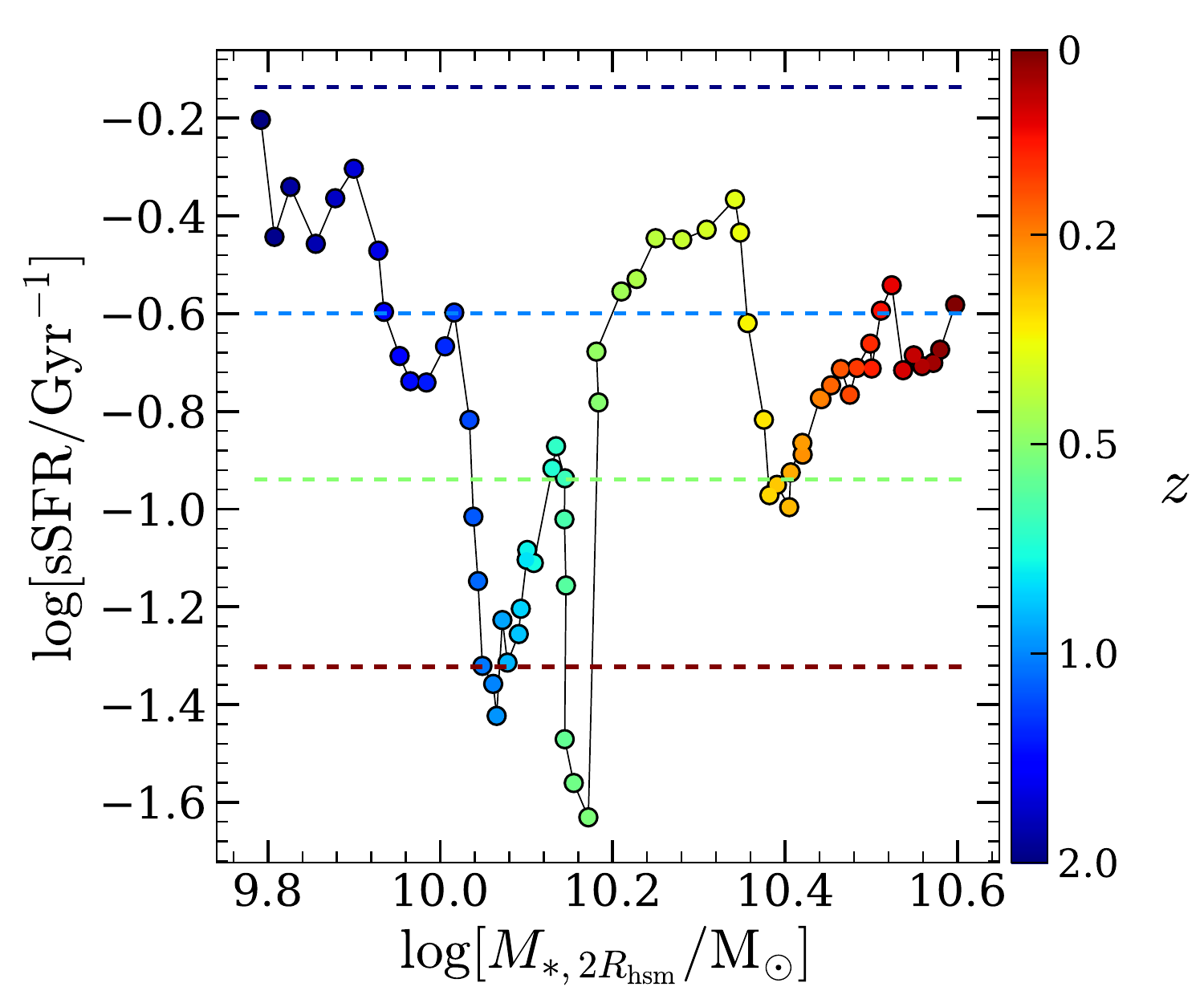} 
\caption{The $\log\,\mathrm{sSFR}-\log\,M_{\ast}$ evolutionary track for a present-day normal massive disc galaxy (ID-434303). The colors indicate the snapshots (redshifts) of the galaxy and the dashed lines indicate the lower boundary in $\log\,\mathrm{sSFR}$ for the main sequence galaxies at different redshifts (indicated by colors), below which galaxies are regarded as quenched. The lower boundary is calculated as the 16th percentile of the sSFR for the galaxies with $\log\,M_{\ast}/\mathrm{M_{\odot}}<10.5$ at different redshifts.}
\label{fig:episodicSF}
\end{figure}

In Fig.~\ref{fig:episodicSF}, we show the star formation history for one example galaxy. The dashed lines in the figure indicate the lower boundary in logarithm of the specific star formation rate (sSFR, defined as the SFR to stellar mass ratio within $2\,R_{\rm hsm}$) for the main sequence galaxies at different redshifts (indicated by colors), below which galaxies are regarded as quenched. As can be seen, during its evolution history, this galaxy had move to and fro between star-forming and quenched states multiple times, exhibiting a typical episodic star-formation pattern among star-forming galaxies. It is worth noting that \citet{Guo21EpisodicHISFQ} made use of the stacked HI spectra from both the Arecibo Fast Legacy ALFA Survey (ALFALFA; \citealt{Haynes18ALFALFAHI, Jones18ALFALFAHI}) and the Sloan Digital Sky Survey (\citealt{Albareti17SDSSDR13}) and also discovered an episodic evolution path for the HI gas in the galaxy, which suggests that a galaxy may go through cycles of gas accretion, gas compaction and quenching before being rejuvenated and starting a new episode of star formation, fully in line with this study. We note that, such an episodic behaviour through star-forming and quiescent phases only shows up in the SFR-mass diagram (as presented in Fig.~\ref{fig:episodicSF}), but not in the color-mass diagram (as figure 10 in \citet{Faber07RedSequenceColorPath} for the red-sequence galaxies). This is because the galaxy color does not necessarily become significantly redder during the short quenched phase, transitioning into the next star-forming episode.

\section{Motion of the cold circumgalactic gas: tangential inflow}
\label{sec:cold}
As already demonstrated in the previous section, the CGM gas motion strongly affects a galaxy's star formation activities. In this section, we directly show the cold gas configuration and their kinematic status.

\subsection{The cold gas' in-spiral pattern}

\begin{figure*}
\centering
\includegraphics[width=2\columnwidth]{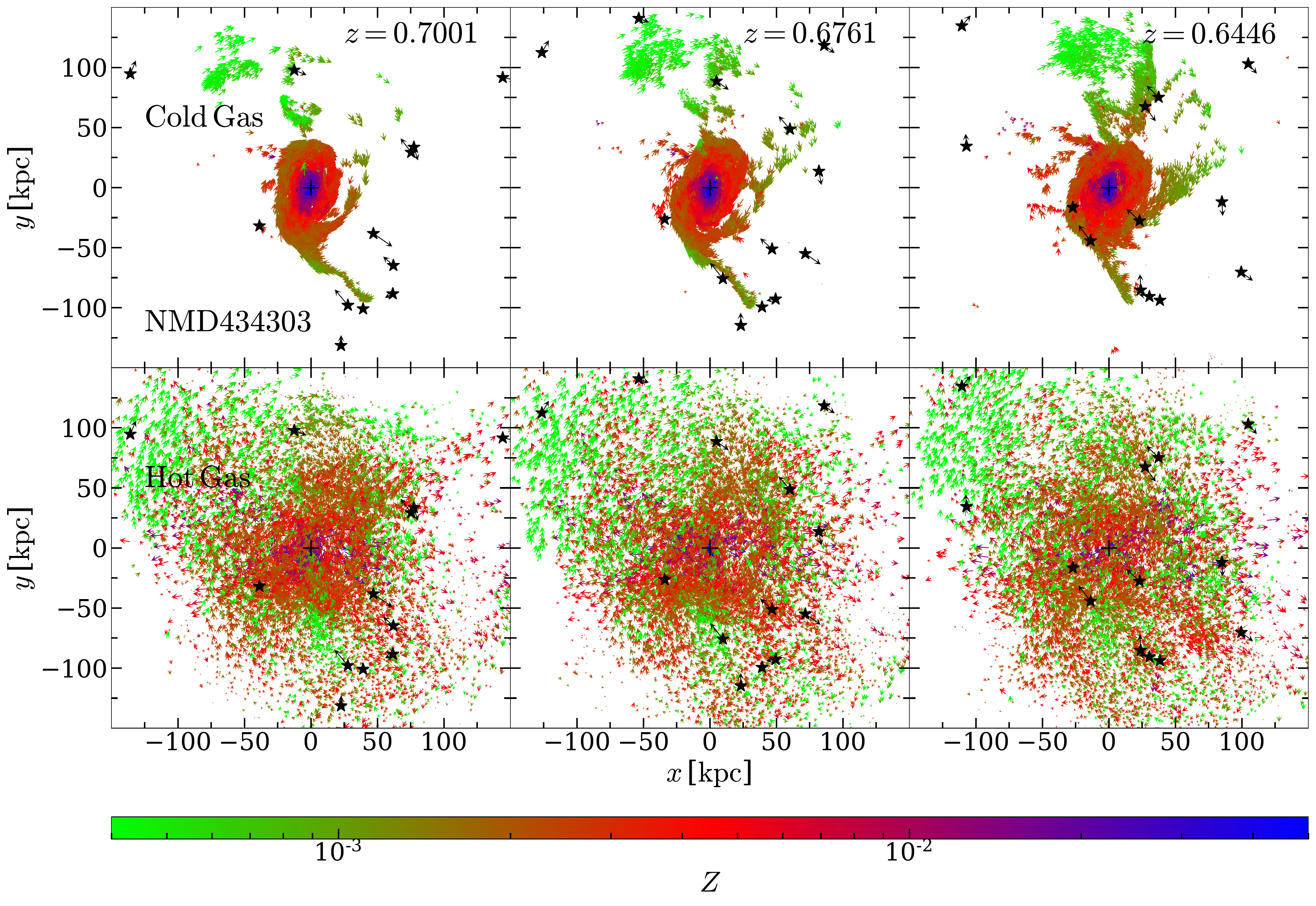}
\caption{Configurations of both the cold (top) and the hot (bottom) circumgalactic gas of a present-day normal massive disc galaxy (ID-434303) when it was at three redshifts. The galaxy is projected along the z-axis of the simulation box. In each panel, the thin vectors indicate the velocities on the $x-y$ plane of the gas cells that are gravitationally-bounded to the central galaxy. The colors and the lengths of the vectors indicate the metallicity and the speeds of the gas cells. The black stars represent the satellite galaxies that have $M_{\rm tot}>10^9\,\mathrm{M_{\odot}}$ and are located around the central galaxy. The black arrows associated with the satellite galaxies (black stars) represent the velocity of the satellites on the $x-y$ plane. For visual purposes, the normalization of the galaxy velocity vectors are 1.5 times that of the gas velocity vectors. The black plus symbol at the center represents the central galaxy.}
\label{fig:configuration of gas}
\end{figure*}

The upper panels of Fig.\,\ref{fig:configuration of gas} show the distribution and motion of the cold circumgalactic gas by presenting the projected velocity vector field on top of the spatial distribution. As can be seen, the cold gas almost ``clusters'' together from the center to few tens of kpc, displaying an in-spiral pattern around the potential minimum (the black plus symbol). The color-coded metallicity distribution indicates decreasing metallicity from inside to outside, which is easily understood as the gas pollution always happens from the inner region due to star-forming activities. 

In Section~\ref{sec:MergingEnvironment}, we study possible origins of the cold circumgalactic gas. As it is carrying angular momentum inherited from galaxy mergers/interaction, the infalling CGM gas will naturally exhibit an in-spiral pattern. With this new supply feeding the galaxy, star formation will turn on. Stars coming into being in this environment will naturally inherit such (primordial orbital) angular momentum so that the stellar disc may largely co-rotate with the outer circumgalactic gas, as well as with the orbital motion of neighbouring galaxies on large scales. We note that such a co-rotating kinematic pattern is observed for the simulated galaxies and is a natural consequence of angular momentum modulation by the environment upon the CGM, which, as shall be seen in Paper II, has a larger impact on present-day quenched galaxies than on star-forming discs. We will give a detailed discussion of such coherent kinematics and its observational signatures in Paper II.

\subsection{The cold gas' tangential and infalling kinematics}

\begin{figure*}
\centering
\includegraphics[width=1\columnwidth]{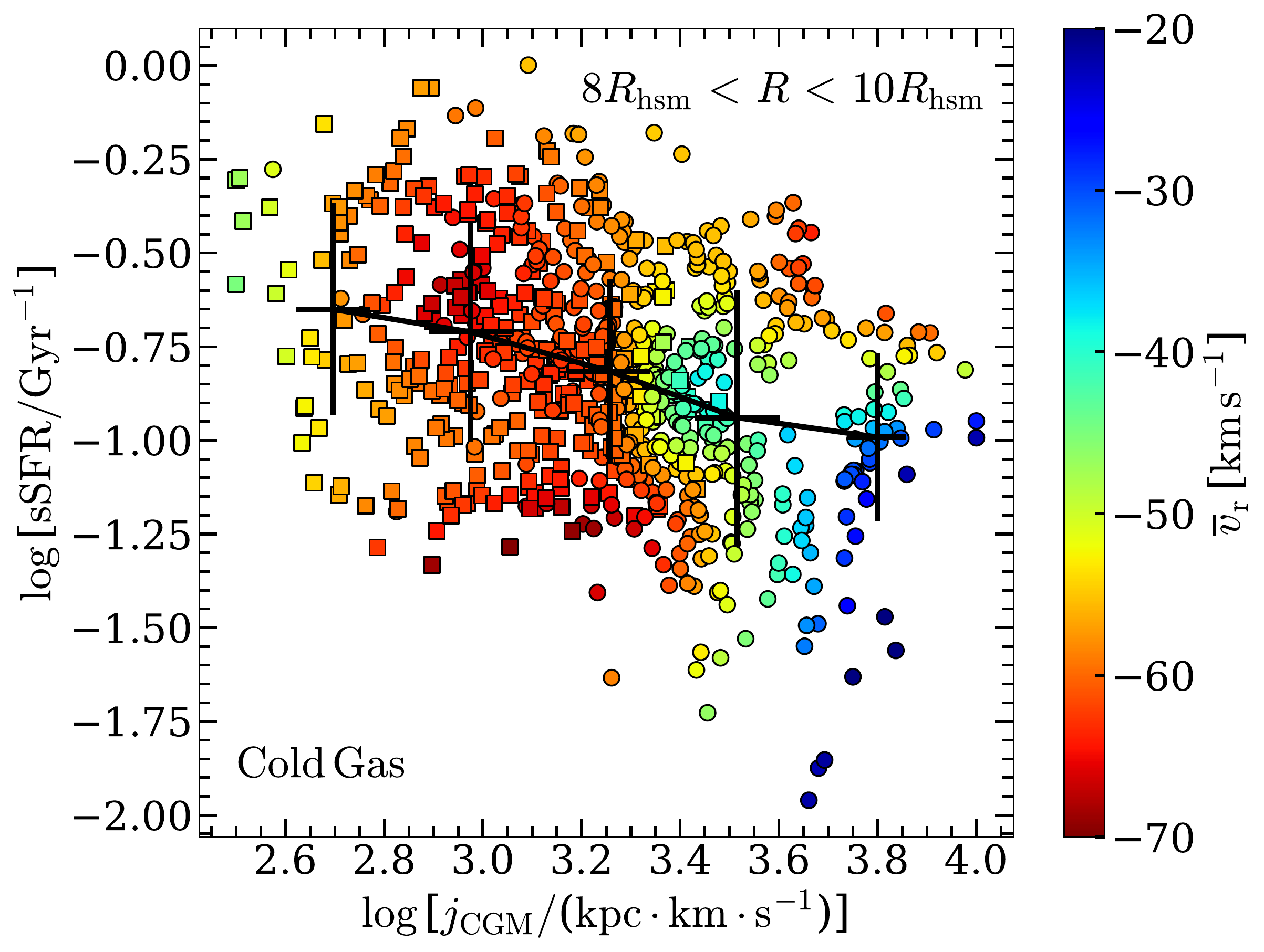}
\includegraphics[width=0.98\columnwidth]{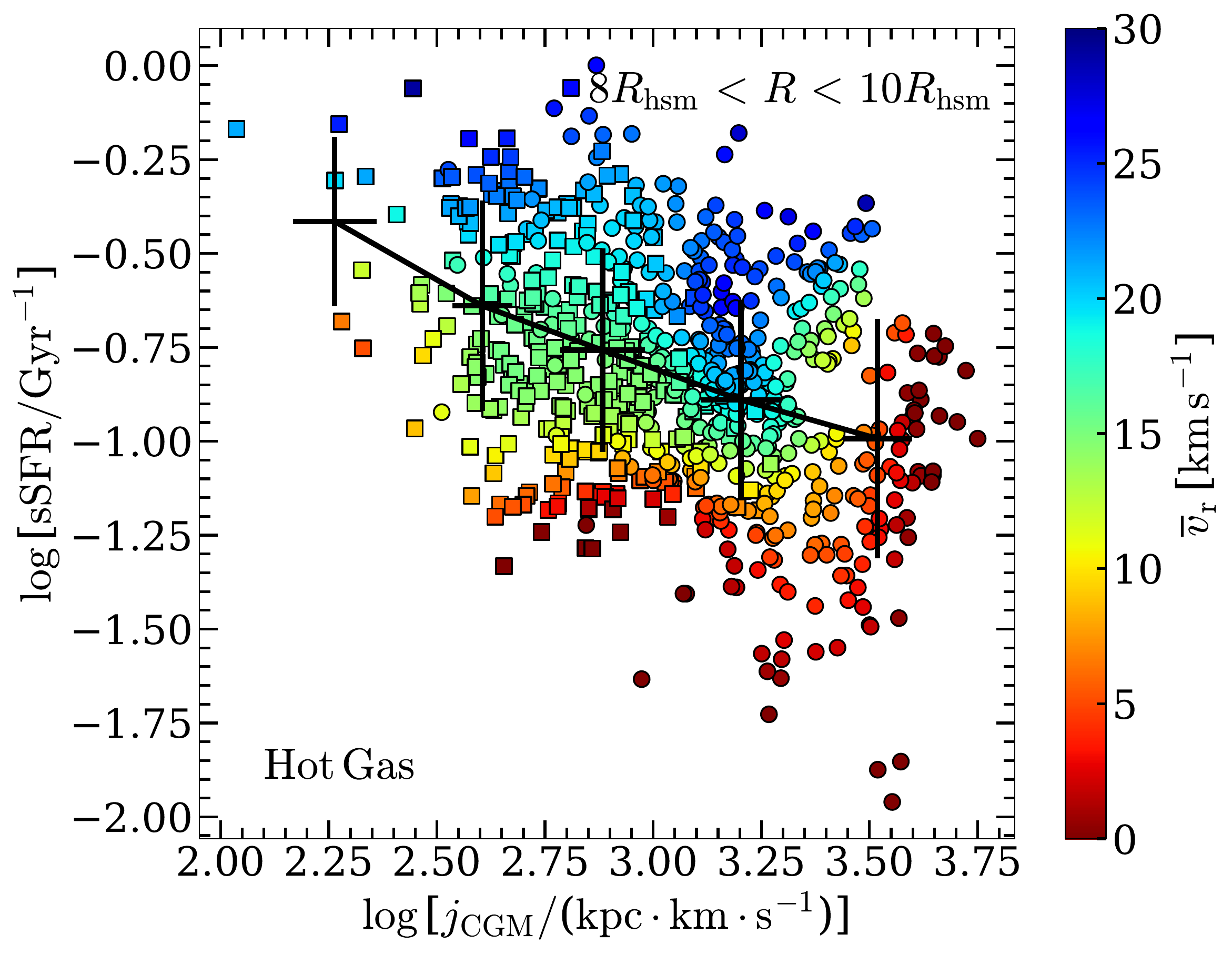}
\caption{The correlation between the specific star-formation rate (sSFR) within $2R_{\rm hsm}$ and the specific angular momentum ($j_{\rm CGM}$) of the cold (left) and hot (right) CGM in a shell between $8R_{\rm hsm}$ and $10R_{\rm hsm}$, color-coded by smoothed radial velocities for 15 galaxies and their progenitors since $z=1$ (circles for NMDs and squares for NDs, each scatter represents one galaxy at one snapshot). Black dots and error bars indicate the averages and $1\sigma$ regions of the corresponding bins. The color-coded distributions of $\overline{v}_{\rm r}$ are smoothed using the {\sc python} implementation of two-dimensional Locally Weighted Regression method \citep{Cleveland&Devlin} (for details see \citealt{Cappellari_et_al_2013b}).}
\label{fig:SFRCGMspin}
\end{figure*}

In this section, we show the kinematic properties of the cold circumgalactic gas. To do so, we introduce an averaged radial velocity $\overline{v}_{\rm r}$, a specific angular momentum $j_{\rm CGM}$ and a spherical anisotropy parameter $\beta \equiv 1-\frac{\sigma_{\theta}^2+\sigma_{\phi}^2}{2\sigma_{\rm r}^2}$, where $\sigma_{\rm r}$, $\sigma_{\theta}$, $\sigma_{\phi}$ represent radial, elevation and azimuthal velocity dispersion, respectively (see \citealt{Binney_and_Tremaine(1987)} for definition).  The left panel of Fig.\,\ref{fig:SFRCGMspin} shows the distribution of the specific star formation rate (sSFR) (evaluated within 2$R_{\rm hsm}$) as a function of the specific angular momentum $j_{\rm CGM}$ of the cold circumgalactic gas within a shell between 8$R_{\rm hsm}$ and 10$R_{\rm hsm}$, color-coded by the mean radial velocity $\overline{v}_{\rm r}$ (of the cold CGM) measured in this shell. The plotted data points represent 15 traced galaxies and their primary progenitors since $z=1$. In the figure, the color-coded distributions of $\overline{v}_{\rm r}$ are smoothed using the {\sc python} implementation\footnote{The software is available from \url{https://pypi.org/project/loess/}} (see details in \citealt{Cappellari_et_al_2013b}) of the two-dimensional Locally Weighted Regression method \citep{Cleveland&Devlin}. 

We first notice that the average $\overline{v}_{\rm r}$ of the gas within the investigated radial range is always negative among different galaxies, implying a universal inflow motion of the cold gas. Also, it can be clearly seen that the specific angular momentum of the cold circumgalactic gas exhibits an anti-correlation with SFR: the lower (higher) the angular momenta, the faster (slower) the gas flows into the galaxy center. Such a trend is observed in both lower-mass and more massive galaxy samples, separately. A higher in-fall speed would further imply a higher efficiency of the cold circumgalactic gas in feeding the galaxy centre, triggering more active star formation. In Paper II, we extrapolate this result to an even higher $j_{\rm CGM}$ regime, which is more common for present-day quenched elliptical galaxies. High CGM angular momenta can effectively prevent the gas from infalling. Thus when a galaxy with a high CGM angular momentum has exhausted its existing central gas reservoir due to star formation and feedback activities, it will remain quenched -- a fate in contrast to an episodic star formation history as for present-day star-forming disc galaxies in this study.

It is worth noting that the above-presented anti-correlation between galaxies' sSFRs and the specific angular momenta of the cold circumgalactic gas could have been a consequence of mass dependencies of the two properties such that the more massive a galaxy is, the less sSFR it has, and independently, the higher CGM angular momentum it associates with. However, a higher mass would result in a larger infalling speed, yielding a pattern in $\overline{v}_{\rm r}$ which increases with higher $j_{\rm CGM}$, the opposite of the trend presented in Fig.\,\ref{fig:SFRCGMspin}. Therefore we expect the anti-correlation exhibited here is more of an angular-momentum argument rather than a consequence of mass dependencies. In paper II, with a larger galaxy sample we explicitly demonstrate that such an anti-correlation is not a mass-driven consequence, indicating the crucial connection between star formation activity and the spin of the CGM.

\begin{figure}
\centering
\includegraphics[width=1\columnwidth]{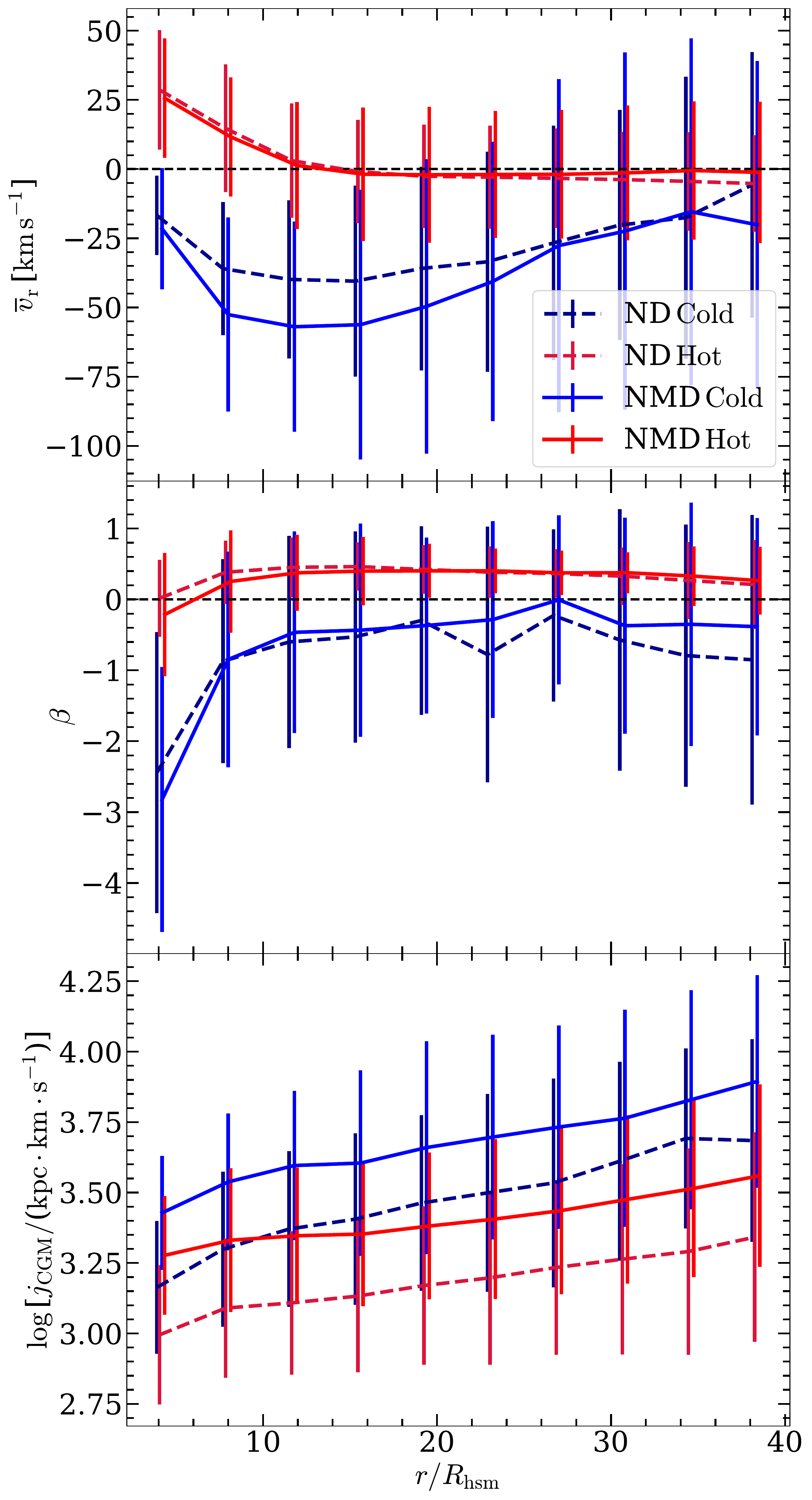}
\caption{Radial profiles of $\overline{v}_{\rm r}$, $\beta$ and $j_{\rm CGM}$ of CGM gas from top to bottom. These parameters are all estimated for the entire $z=0$ samples. In each panel, the blue and the red curves represent cold and hot CGM gas, respectively (In order to distinguish the error bars, we set the NDs with darker red and blue). Normal discs (NDs) and normal massive discs (NMDs) are indicated by dashed and solid curves, with the error bars showing the $1\sigma$ region of the investigated parameters. The black dashed lines mark $\overline{v}_{\rm r}=0$ (top) and $\beta=0$ (middle).}
\label{fig:vr_beta_ang_profile}
\end{figure}

To see how the CGM gas kinematics scales with radius, we show, in Fig.\,\ref{fig:vr_beta_ang_profile}, the radial profiles of three parameters: the averaged radial velocity $\overline{v}_{\rm r}$, the anisotropy parameter $\beta$, and the specific angular momentum $j_{\rm CGM}$ of the CGM gas for our $z=0$ galaxy sample. The two galaxy samples at different mass scales (NDs versus NMDs) show similar radial profiles, only that more massive discs exhibit larger infalling speed and higher CGM angular momenta. Again $\overline{v}_{\rm r}$ of the cold gas is on average always negative, suggesting an overall infalling motion on all scales. One interesting thing is that as the cold gas flows in, $|\overline{v}_{\rm r}|$ increases towards smaller distances (i.e., the inflow becoming faster due to gravity) until $R\sim 10-15\,R_{\rm hsm}$, at which point it turns around and begins to decrease. This is direct evidence that the cold gas encounters a hydro-resistance from the out-flowing hot gas (see next Section). The averaged $\beta$ profile of the cold gas is also always negative, implying a tangential motion (rotational feature) across a large range of distances. Together, the cold gas exhibits in-spiralling kinematics with a configuration as demonstrated in Fig.\,\ref{fig:configuration of gas}.

\section{Motion of the hot circumgalactic gas: radial outflow}
\label{sec:hot}

\subsection{The hot gas' distribution is rather smooth}

The pristine gas that a massive galaxy acquires through hot-mode accretion cannot cool rapidly after being shock-heated to the virial temperature (e.g., \citealt{Dekel_et_al_2009}). Within the galaxy, AGN and stellar feedback activities play a dominant role in heating and lifting up the inner (metally-enriched) gas to larger distances well into the CGM regime.  Although the hot circumgalactic gas does not directly feed central star formation, it can ultimately fuel the central cold gas reservoir through cooling (at larger distances) and can influence the motion of the cold gas. Quite different from the cold gas, the distribution of the hot circumgalactic gas, as can be seen from the bottom panels in Fig.\,\ref{fig:configuration of gas}, is rather smooth extending to very large scales ($\sim 100\,\mathrm{kpc}$). Again as the radius decreases, the gas-phase metallicity increases due to central star forming activities. 

\subsection{The hot gas' radial and outward kinematics}

The right panel of Fig.~\ref{fig:SFRCGMspin} shows the distribution of the sSFR as a function of $j_{\rm CGM}$ for the hot circumgalactic gas in a shell of $8\,R_{\rm hsm}-10\,R_{\rm hsm}$. We can see that although the hot gas does not participate in star formation, sSFR still anti-correlates with the hot gas' specific angular momenta. But quite different from the cold counterpart, instead of markedly changing with respect to $j_{\rm CGM}$ (horizontally), the $\overline{v}_{\rm r}$ distribution of the hot gas varies strongly depending on the sSFR (vertically): the higher the sSFR is, the faster the hot gas flows outwards, unveiling the effect of feedback.

In terms of gas dynamics as a function of radius, the hot circumgalactic gas possesses lower angular momentum and exhibits more radially out-going kinematics, in comparison to its cold gas counterpart. The results are shown in Fig.\,\ref{fig:vr_beta_ang_profile}. Specifically, the average $\beta$ profile of the hot gas is largely positive, indicating its radial motion. At smaller radii ($R< 15\,R_{\rm hsm}$), the out-flowing hot gas moves with ever increasing velocities near the galactic centre due to stronger and stronger repulsive feedback effects. At larger radii beyond $R\sim 15\,R_{\rm hsm}$, the mean radial velocities become nearly zero, implying a certain dynamical equilibrium in the radial motion between the out-going (due to feedback) and infalling (due to accretion) hot gases at these distances. As noted already, such a transition distance of $\sim 15R_{\rm hsm}$ ($45-75\,\rm kpc$) is where the cold CGM gas starts reducing its infalling speed as it interacts with the ejected hot gas from the galactic centre. This radius can be viewed as an approximate boundary to a region within which feedback can have a strong influence.

\section{Galaxy interactions on larger scales trigger new episodes of star formation}
\label{sec:MergingEnvironment}
In Section~\ref{sec:breathing}, we have shown the entangled fates between the ``breathing'' circumgalactic gas and the episodic star formation. A crucial condition for this to happen in a repetitive and sustainable fashion relies on a galaxy's large-scale environment, in which galaxy mergers and fly-by interactions continuously provide effective channels to send cold gas from large scales down to small radii inside the galaxy, where the star-forming interstellar medium gets replenished. In this section we present key evolutionary stages of individual galaxies in their large-scale environments (Section~\ref{sec:evd1} and \ref{sec:evd2}) and an argument from the metallicity perspective to identify the different origins of the cold and hot circumgalactic gases (Section~\ref{sec:evd3}), all indicating the crucial role played by galaxy interactions in triggering new epochs of star formation. We refer the reader to Paper II of this series, where using a statistical galaxy sample, we show that the large-scale environment strongly affects and regulates the angular momentum of the circumgalactic gas, which further influences star-forming and quenching activities inside the galaxy. It is worth noting that the large-scale environment serves as a necessary condition for episodic star formation to happen; however, the time scale of each episode is essentially governed by the consumption rate of the galaxy's cold gas reservoir, which is strongly regulated by AGN and stellar feedback activities (see Section~\ref{sec:hot} for more detailed discussion).

\subsection{Evidence $\#1$: Concurrences between galaxy interactions and new star-formation episodes}
\label{sec:evd1}

There have been many works in the literature investigating the effects of galaxy mergers and interactions on star formation activities (e.g., \citealt{Barnes_et_al.(1992),Barnes_et_al.(1991),Barnes_et_al.(1996),Mihos_et_al.(1996), Barnes_et_al.(2002),Naab_et_al.(2003),Bournaud_et_al.(2005),Bournaud_et_al.(2007),Johansson_et_al.(2009a),Tacchella_et_al.(2016),Rodriguez-Gomez_et_al.(2017)}). In this subsection, we show specific evidence in the evolution of individual galaxies from the simulation.

In Fig.~\ref{fig:example_track_ND} and \ref{fig:example_track_NMD} we present, for eight selected galaxies in our traced sample, the redshift evolution of gas fraction, metallicity, and star-forming phase-space, together with the merger histories of the galaxies. We explicitly look for typical features which show connections between galaxy interactions and active star-forming epochs.
The top row in each panel shows again the ``anti-correlation'' between the central gas fraction and gas metallicity, as discussed in Section~\ref{sec:breathing}. From the third row, we see that many new epochs of star formation coincide with registered merger events, as indicated by the vertical lines throughout the four rows in each panel. This, for example, can be seen in the event that happened at $z\sim 0.6-0.7$ for galaxy ID-601507, as a clear showcase. While a series of successive merging events can often be seen overlapping with continuous star-forming epochs, e.g., in recent histories since $z\sim0.4$ for galaxies ID-529270, ID-483977 and ID-496369. This suggests that galaxy mergers may effectively drive the cold circumgalactic gas down to fuel the galaxy (central) star formation. This cold gas is often of a lower-metallicity, which is why new epochs of star formation often start from low metallicity states (\citealt{Bustamante18MergerInduceLowMetallicity}). We mention in passing that the lower-metallicity branch seen in the star-forming phase-space diagram corresponds to star-forming sites at larger galacto-centric distances; while the higher-metallicity one corresponds to the galactic central regions (i.e. bulges). 

We note that not all galaxy interactions are in the form of galaxy mergers, where the accreted galaxy eventually joins the host galaxy and the event is registered as a merger. It is interesting to notice that some of the star-forming episodes are not associated with any registered merging events. However, as can be seen from the second row, these star-forming epochs happen to coincide with enhanced environmental/neighbour galaxy number densities, which are counted either within $R_{200}$ (denoted as $n_{R_{200}}$), or within a sphere which contains all group member galaxies that are more massive than 1 per cent of the central/host galaxy (denoted as $n_{\rm loc}$). A large number density indicates an environment with a higher chance of galaxy interaction (either merger or fly-by). Such concurrences between large number densities and new episodes of star formation can be clearly seen for galaxy-ID 527857 around $z \sim 0.3-0.8$ and for galaxy ID-503826 since $z=1$.

\subsection{Evidence $\#2$: cold circumgalactic gas associated with infalling galaxies}
\label{sec:evd2}

\begin{figure*}
\includegraphics[width=2\columnwidth]{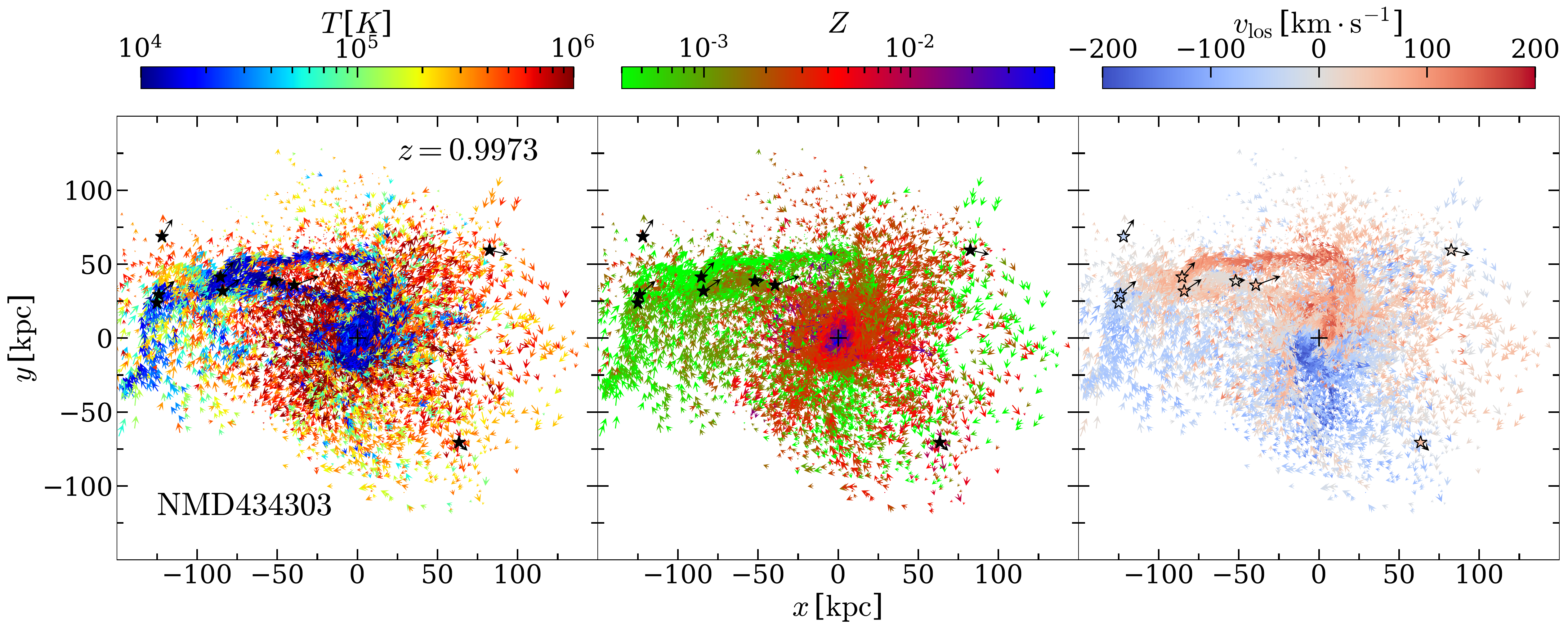}
\caption{Gas configurations of a present-day normal massive disc galaxy (ID-434303) when it was at $z\sim 1$, color-coded by the gas temperature (left), the metallicity (middle) and the line-of-sight velocity (right). These plots are projected along the z-axis of the simulation box. The symbols are the same as Fig.~\ref{fig:configuration of gas}.} 
\label{fig:NMD_434303}
\end{figure*}

\begin{figure*}
\includegraphics[width=2\columnwidth]{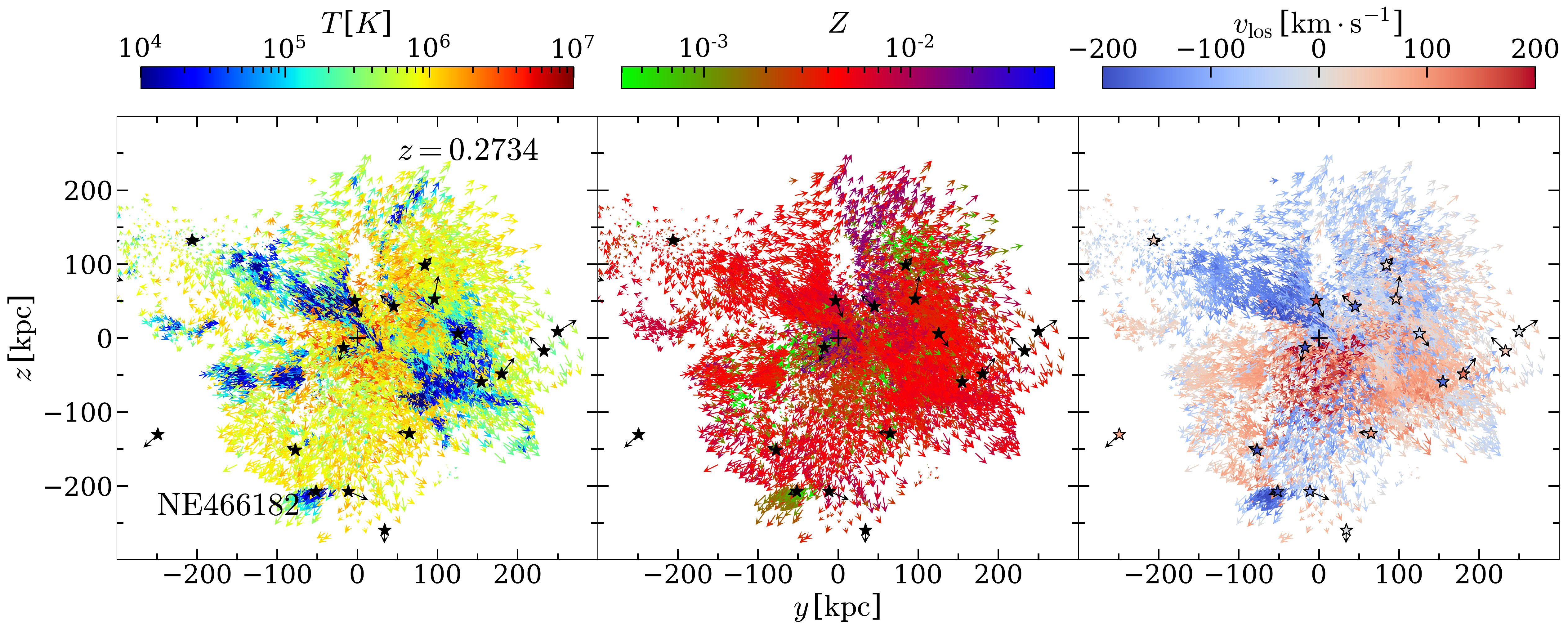}
\includegraphics[width=2\columnwidth]{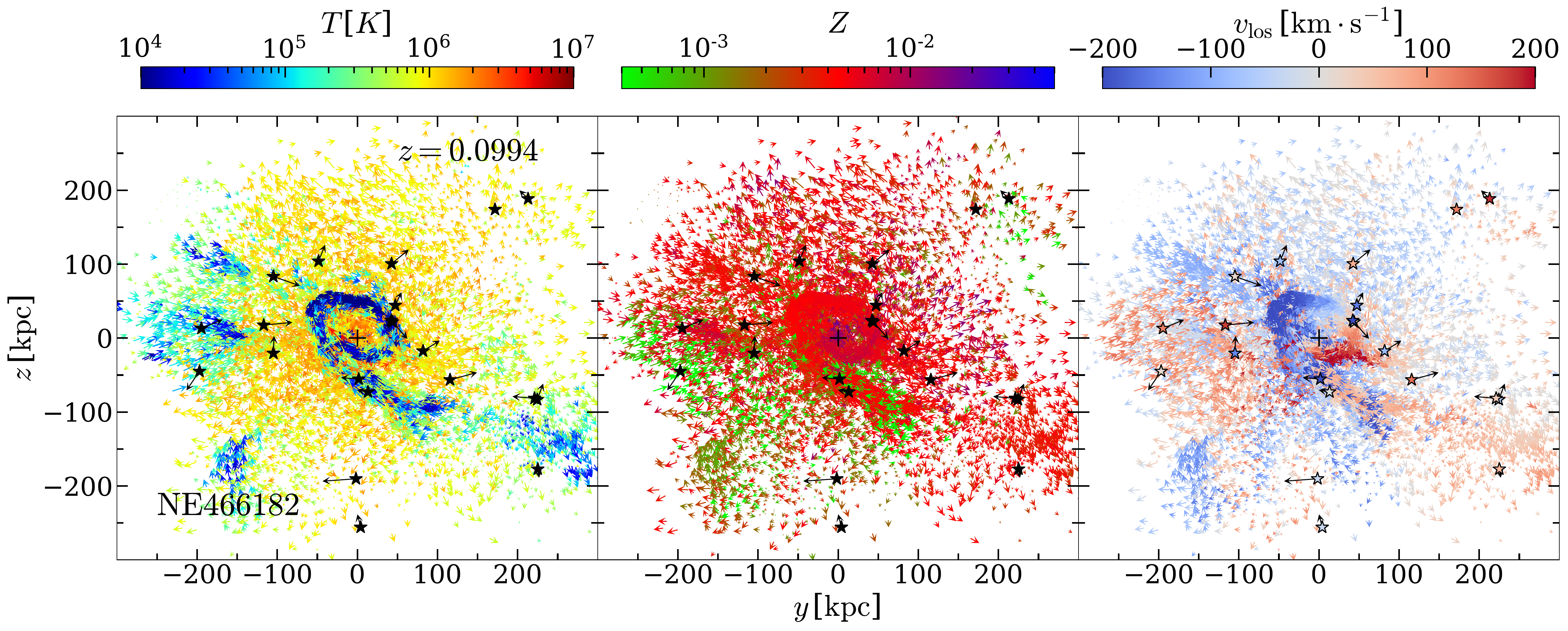}
\caption{Gas configurations of a present-day early-type galaxy (ID-466182) when it was at $z\sim 0.27$ (top) and $z\sim 0.1$ (bottom). These plots are projected along the x-axis of the simulation box. The symbols are the same as Fig.~\ref{fig:configuration of gas}. }
\label{fig:NE_466182}
\end{figure*}

A second piece of evidence comes from infalling cold gas associated with merging-in/flying-by galaxies. We emphasize that this is in line with the findings of Paper II, which demonstrates that the cold circumgalactic gas inherits angular momentum from the large-scale environment through mergers and fly-by interactions. Here we provide online (\url{https://astro.tsinghua.edu.cn/~dxu/HTM/Science/MetTemEvolutionExample_WXL21.htm}) two animations for motions of the CGM gas and of the satellite galaxies that belong to two example galaxies since $z=4$. The first example galaxy (ID-434303) is a normal massive star-forming disc galaxy at $z=0$. As can be seen in the animation, a group of incoming galaxies at $z\sim 1.8$ appear from the 2 o'clock position in the animation domain. Streams of cold and lower-metallicity gas follow these galaxies closely, as they spiral around the host galaxy. During this time, the gas drops down from as far as $\sim 100\,\rm kpc$ towards the host center in an in-spiral fashion. This feature is particularly strong around $z\sim 1$. We present the gas motion and properties in this galaxy at this exact snapshot in Fig.~\ref{fig:NMD_434303}, where each arrow indicates the velocity of a gas cell, color-coded by the gas temperature (left), metallicity (middle) and the line-of-sight velocity (right). We note that all plotted gas elements are already gravitationally-bound to the central host galaxy at each snapshot. We did not trace the gas cells to identify a physical origin of the gas regarding whether they are stripped from the satellite galaxies or originated in the host galaxy. However, we confirm that these gas particles are associated with the incoming galaxies in their 3D position, gas metallicity as well as line of sight velocity.

The next example galaxy (ID-466182) is a normal quenched elliptical galaxy at $z=0$ (from \citealt{Lu2021ColdElliptical}). This galaxy has completely ceased its star-formation since just below $z \sim 0.4$ and had not experienced any registered merger activities since then until $z \sim 0.1$. In Fig.~\ref{fig:NE_466182}, we show the gas velocity fields, color-coded by temperature, metallicity, and line-of-sight velocity for this galaxy at two different redshifts ($z=0.2734$ and $z=0.0994$). At both redshifts, the cold gas is not obviously seen to be associated with any particular incoming galaxy. Such cold streams are in fact very often present during a galaxy's dynamical evolution. Their inflow directions tend to trace the trajectories of fly-by galaxies. The filamentary cold gas at a distance of $\sim 100\,\rm kpc$ also has metallicity as high as that in the inner region of the galaxy ($\sim 0.1-1\,Z_{\odot}$). We argue that the formation of such cold and metal-rich infalling gas can easily be achieved: the metal-enriched circumgalactic gas that had been polluted and flown out to larger radii due to previous epochs of star formation and AGN feedback can start to cool efficiently as a result of shock compression, as neighbour galaxies fly by. In \citet{Cai17LyANebAtZ2}, the authors present observational evidence that both shock-induced collisional ionization and photo-ionization are required in order to produce efficient emission-line cooling of metals observed at large distances; the cooled and metal-enriched circumgalactic gas then inflows back to the galaxy in a spiral fashion.

Through this second example, we would also like to demonstrate that in-spiraling kinematics of the cold circumgalactic gas is not a unique feature solely in star-forming disc galaxies. It can also exist in quenched elliptical galaxies. As shown in the bottom panel of Fig.~\ref{fig:NE_466182}, such a pattern in the cold circumgalactic gas emerges gradually from $z \sim 0.2$, exhibiting itself as an in-spiral structure right before a registered merging event that happens at $z\sim 0.1$. Interestingly, this gas inflow did not lead to notable star formation in this system. As studied in Paper II of this series, this has to do with the large angular momentum associated with the revolving gas, which prevented it from efficiently flowing into the galaxy center to trigger new star formation.

\subsection{Evidence $\#3$: gas metallicity distributions at large scales}
\label{sec:evd3}

\begin{figure}
\includegraphics[width=1\columnwidth]{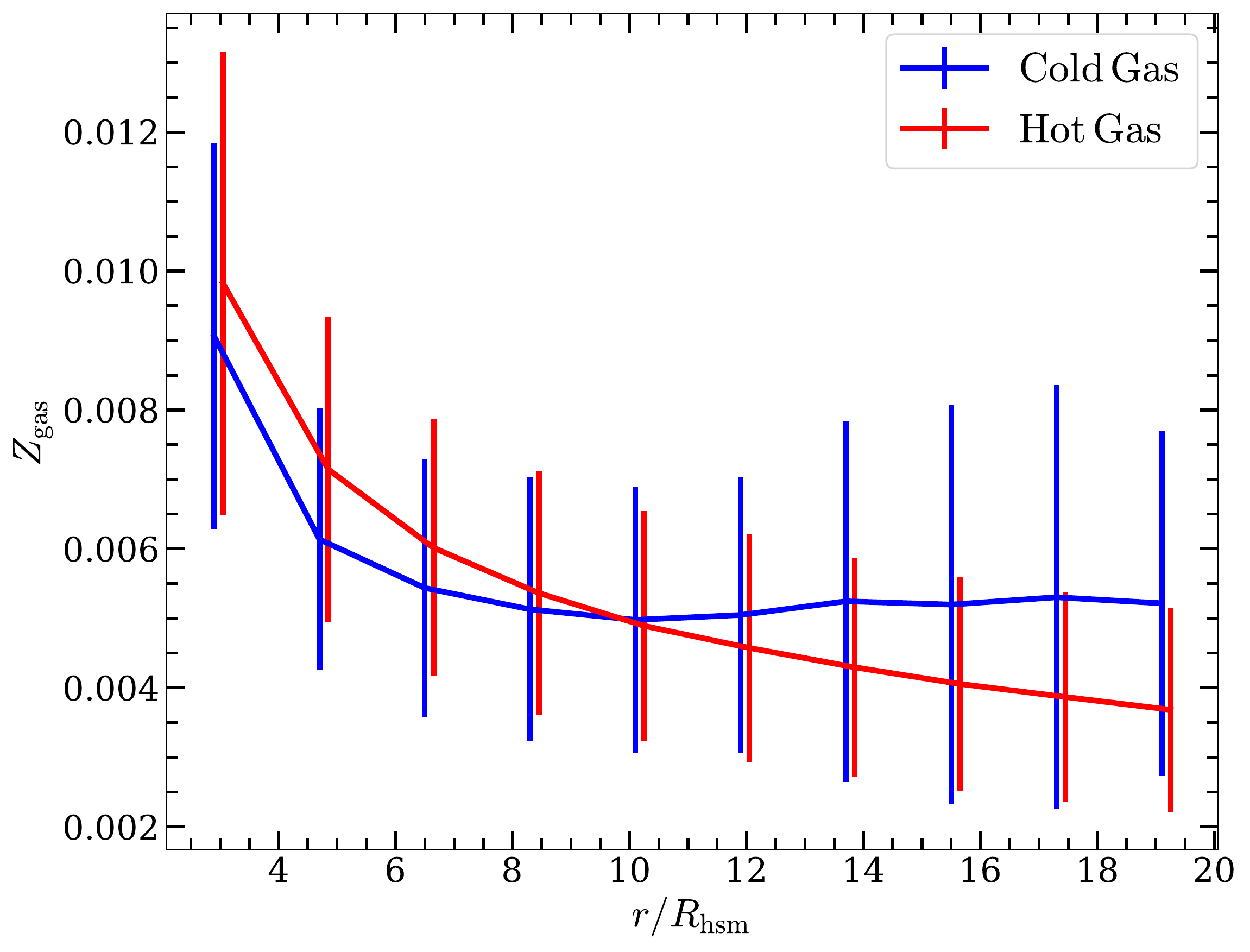}
\caption{The radial profiles of the gas-phase metallicity for all 15 selected galaxies at $z=0$. The blue and red curves denote the results for cold and hot CGM gas, respectively, with the error bars indicating the $1\sigma$ region. }
\label{fig:metal_profile}
\end{figure}

The third tentative piece of evidence that links the origin of the cold circumgalactic gas with merger activities is that at larger distances beyond $\sim 10 R_{\rm hsm}$, the cold gas metallicity becomes markedly higher than the hot gas metallicity. The opposite is true at distances within. This can be seen from Fig.\,\ref{fig:metal_profile}, which presents the averaged radial profiles of the cold- and hot-gas metallicities for our selected star-forming galaxy samples at $z=0$. At these large distances, the average radial motion of the hot circumgalactic gas becomes less directly affected by the central feedback activities (see Section~\ref{sec:cold}). The poorer metallicity (in comparison to that of the cold CGM gas at the same distances) reveals its pristine IGM origin, which is different from the heavily enriched metal environment at smaller radii near the galactic centre. While the metal-richer cold gas seen at these distances is more likely to be stripped from interacting galaxies, with time it becomes gravitationally bound to the host galaxy, and eventually spirals in to feed the central cold gas reservoir.

\section{Observational consequence - episodic star formation histories}
\label{sec:observation}

\begin{figure*}
\includegraphics[width=1\columnwidth]{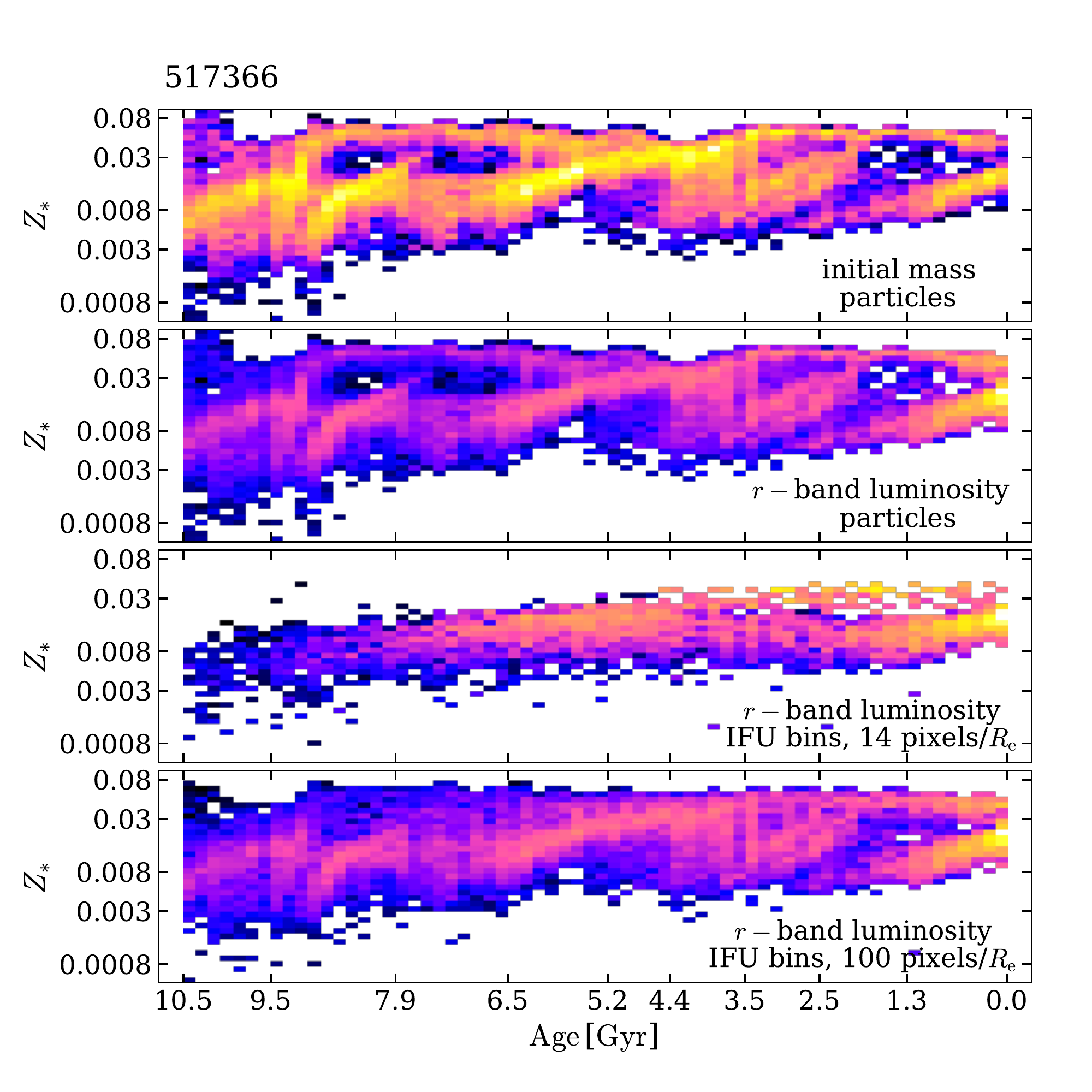}
\includegraphics[width=1\columnwidth]{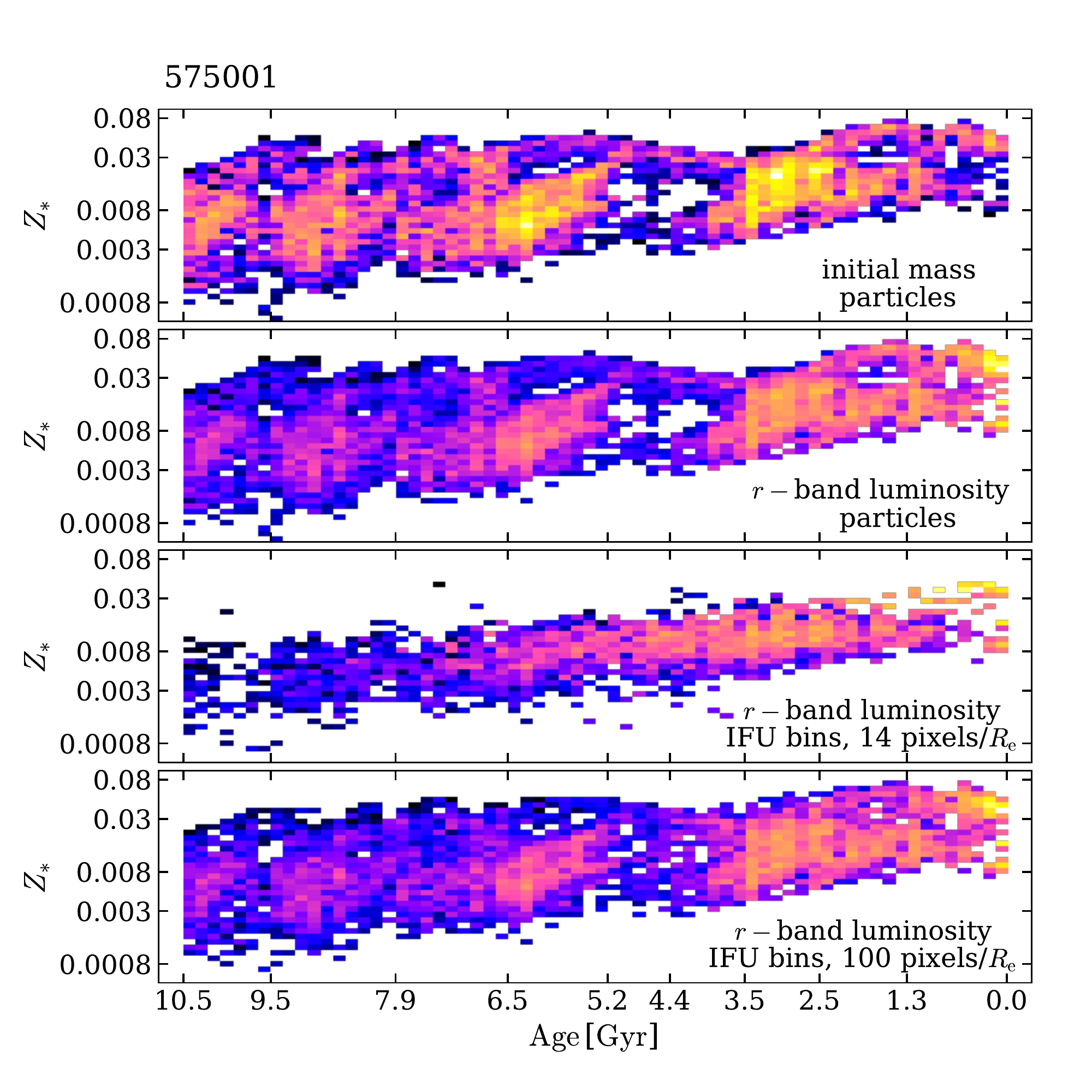}
\caption{The age-metallicity phase space distributions for two example galaxies (left: ID-517366; right: ID-575001). In each panel, the colors indicate the intensities (of mass or luminosity) of ``stars'' that are born with a given metallicity and at a given redshift (age). From top to bottom, the intensity of each bin in the age-metallicity phase space is calculated as: \textbf{(1)} the total initial mass (the mass at birth) of particles in this bin; \textbf{(2)} the total present-day $r-$band luminosity of particles in this bin; \textbf{(3)} the total present-day $r-$band luminosity of corresponding pixels of the (luminosity-weighted) age and metallicity maps from the face-on view of the galaxy; a finite spatial resolution of 14 pixels per $R_{\rm hsm}$ (the median resolution for MaNGA galaxies) is adopted; and \textbf{(4)} the total $r-$band luminosity of corresponding pixels in the age and metallicity maps as in the case for \textbf{(3)}, while the spatial resolution is 100 pixels per $R_{\rm hsm}$.} 
\label{fig:observation prediction}
\end{figure*}

We have now clearly demonstrated that the large-scale environment influences the central star formation activities by sending the cold circumgalactic gas from large distances down to small radii inside a galaxy. In combination with AGN and stellar feedback activities, together they result in characteristic episodic star formation histories. Observationally, with a large sample of a million main-sequence turn-off stars from the LAMOST spectroscopic survey, \citet{Xiang17LAMOSTAgeMetTwoEpisode} suggested the existence of more than one sequence in the age-metallicity distribution of nearby disc stars, including at least an older sequence (with stellar age $>$ 8\,Gyr) in the inner disc, and a younger sequence  ($<$ 5\,Gyr) in the outer disc (see their section 6.4 and figure 22). Based on a smaller sample but with high-resolution spectroscopic data, \citet{Nissen2020_TwoEpisodesMWstars} detected two sequences in the age-metallicity distribution of nearby solar-type stars from the HARPS spectroscopic program \citep{Mayor03HARPS, Sousa2008HARPSStar}. The clear split of the two sequences indicates the presence of two episodes of gas accretion and star formation, with a quenched phase in between (see figure 17 therein). Such observations have already demonstrated that it is plausible to extract the multiple epochs of star formation that have happened in the history of the Milky Way. In particular, the Gaia mission, in combination with large spectroscopic surveys, has already confirmed several merger/accretion events, as revealed by various satellite streams and kinematic substructures in the Milky Way (see \citealt{Helmi20HistoryMW} and references therein).

For nearby galaxies, this becomes harder as a large sample of individually resolved stars cannot be easily obtained using current telescope facilities. The fossil record of past star formation histories can be deciphered, at best, through spatially resolved stellar populations obtained using integral field unit (IFU) spectroscopic facilities (see \citealt{Peterken20MaNGAStellarPopulation}). In this case,  averaged stellar ages and metallicities are obtained for each and every IFU pixel through stellar population synthesis. The combined data will then be used to construct an age-metallicity phase space distribution of present-day stellar populations across the galaxy. However, a good recovery of the multi-epoch star formation history relies on the spatial resolution of the IFU observation for the galaxy.

We present in Fig.\,\ref{fig:observation prediction} the age-metallicity phase space distributions estimated under different observational conditions for two example galaxies (left: ID-517366; right: ID-575001). The top row shows the stellar mass intensities for stars at birth; i.e., the intensity value of each pixel is calculated by adding initial masses of the stellar particles that sit in a given $Z_{\ast}$-age bin. As can be seen, quiescent phases in between star formation episodes are clearly present. 

The second row presents the luminosity intensities of stars at the present day: the pixel values are obtained by adding the $r$-band luminosities of the stellar particles that sit in a given $Z_{\ast}$-age bin. This row essentially mimics results from resolved stars within the Milky Way. As can be seen, in this case star-forming epochs from as early as 10 Gyr ago in principle can be well-resolved. 

The third and the fourth rows in Fig.\,\ref{fig:observation prediction} also present luminosity intensities of present-day stars. However, in these cases, the pixel values in the age-metallicity phase space are obtained not using individual stellar particles but using the galaxy's (luminosity-weighted) age and metallicity maps of certain finite spatial resolution. For the third row, the spatial resolution is 14 pixel per $R_{\rm hsm}$, equivalent to resolving a galaxy of $R_{\rm hsm}=3\,{\rm kpc}$ at $z=0.02$ with MaNGA-like IFU resolution (0.5$^{\prime\prime}$/pixel). While the fourth row is obtained using age and metallicity maps that are produced assuming a spatial resolution of 100 pixel per $R_{\rm hsm}$ for the galaxy. These last two rows essentially mimic the age-metallicity phase space distributions that one would obtain for typical nearby galaxies using IFU spectroscopic data coming from current and next-generation telescope facilities. As can be seen, the episodic star-forming features are largely washed out for typical MaNGA galaxies (as presented in the third row). We note however, a small fraction of MaNGA galaxies that have larger sizes or are located within 10 Mpc would have a chance to have their multiple star-forming epochs be resolved through IFU spectral synthesis. Next-generation observing facilities such as thirty-meter class telescopes with adaptive optics would be able to reach a higher resolution exceeding 100 pixel per $R_{\rm hsm}$ for a much bigger sample of nearby galaxies to truly carry out extragalactic archaeology.

\section{Discussion and Conclusions}
\label{sec:conclusion}

In this paper, we use star-forming disc galaxy samples from the TNG-100 simulation to study the interconnections between a galaxy's star formation, the circumgalactic gas motion and the larger-scale environment. Our main results are listed below. \\

\begin{enumerate}
\item The circumgalactic gas is ``breathing'' in and out to very large scales, in sync with episodic star forming activities that happen inside the galaxy (see Section~\ref{sec:breathing} and Figs.\,\ref{fig:Evolution_breathing} and \ref{fig:episodicSF} therein). In particular, peaks (troughs) of the star formation rate often associate with precursory decreases (increases) of the cold circumgalactic gas, which stops a further surge (decline) and leads to a fall (rise) of the star formation rate in a later stage (see Fig.\,\ref{fig:relation between cold gas and sfr}). 

\item We divide the CGM gas into a cold phase ($10^{4}\,{\rm K}-2\times 10^{4}\,{\rm K}$) and a hot phase (>$10^{5}\,{\rm K}$). The gas motions in these phases, which are both regulated by the large-scale environment and galactic feedback (see Section~\ref{sec:cold} and \ref{sec:hot}), exhibit, however, completely different spatial configurations (see Fig.\,\ref{fig:configuration of gas}) and kinematic responses (Fig.\,\ref{fig:vr_beta_ang_profile}). The cold circumgalactic gas serves as a feeding source to the central star-forming gas reservoir and carries out a tangentially ($\beta <0$) inwards ($\overline{v}_{\rm r}<0$) motion on all scales. In particular, the lower the angular momentum of the cold gas, the faster it inflows into the galaxy, causing a larger rate of star formation (see the left panel of Fig.\,\ref{fig:SFRCGMspin}). In contrast, the motion of the hot gas is radially ($\beta >0$) outwards ($\overline{v}_{\rm r}>0$). In particular, the more active star formation is, the faster it flows outwards to larger distances, indicating a strong modulation by stellar and AGN feedback activities (see the right panel of Fig.\,\ref{fig:SFRCGMspin}). It is worth noting that despite the fact that the hot circumgalactic gas does not directly fuel star formation, it however, affects the motion of the cold gas through hydrodynamic forces within $\sim10\,R_{\rm hsm}$ and thus lowers the inflow efficiency of the cold gas (the top panel in Fig.\,\ref{fig:vr_beta_ang_profile}). 

\item We present three individual pieces of evidence to illustrate the crucial role played by a galaxy's environmental interactions in triggering new episodes of star formation (see Section~\ref{sec:MergingEnvironment}). Mergers and fly-bys of neighbouring (secondary) galaxies can cause them to lose gas to the primary galaxy, as well as perturb the circumgalactic gas of the primary galaxy (see Fig.\,\ref{fig:NMD_434303} and \ref{fig:NE_466182}). As a consequence, the cold gas falls into the galaxy center in an in-spiral fashion and fuels star formation activities therein. The imprints left by these galaxy interactions are also reflected in a higher metallicity of the cold gas than that of the hot gas (which may originate from virial heating) at a distance further than $\sim 10\,R_{\rm hsm}$ (see Fig.\,\ref{fig:metal_profile}).

\item We predict that episodic star formation histories with clear quiescent phases in between can be extracted in age-metallicity phase space from spatially resolved stellar populations within the Milky Way and/or for nearby galaxies (see Fig.\,\ref{fig:observation prediction}).

\end{enumerate}

The episodic star formation history, as is presented in Fig.\,\ref{fig:Evolution_breathing} and in Appendix~\ref{apd:examples}, is well synced with the pulsating motion of the circumgalactic gas, which is modulated by both the large-scale environment and the central star-forming and feedback activities. Such an episode begins with cold circumgalactic gas feeding the central gas reservoir from large distances, enhancing both star formation and feedback activities inside the galaxy. With time the stellar and AGN feedback come to substantially heat up and blow out the cold gas inside the galaxy, preventing the cold CGM from efficiently feeding the central gas reservoir. This in turn slows down and weakens star forming and feedback activities as the central gas reservoir is being consumed. Eventually the cold circumgalactic gas, as perturbed by galaxy interactions, may once again flow into the galaxy and trigger the next episode of star formation. By repeating the processes above, present-day star-forming disc galaxies have experienced characteristic episodic star formation histories, well synced with the corresponding rhythmic motions of the circumgalactic gas.

It is worth emphasizing that the presence of such an episodic pattern and essentially its time scale are determined by the consumption rate of the galaxy's cold gas reservoir, which is regulated by stellar and AGN feedback activities. In \citet{ElBadry16BreathingFire, ElBadry17JeansFailFeedback}, a ``breathing'' motion was discovered for lower-mass galaxies ($2\times 10^{6}\, \mathrm{M_{\odot}} < M_\ast < 5\times 10^{10}\, \mathrm{M_{\odot}}$) in the FIRE simulation (\citealt{Hopkins14FireSim}), where gas outflows driven by the stellar feedback can generate non-equilibrium phases of the gravitational potential, yielding fluctuations in sSFR and galaxy size, as well as stellar migration. For relatively more massive galaxies as examined in this work, the AGN feedback essentially plays a similar role in enriching, heating and expelling gas out to the CGM regime, regulating star-forming and quenching activities. The role of a large-scale environment is to provide a necessary condition for feeding a star-forming galaxy (and its CGM) not only with a fresh gas supply but also with angular momentum in a sustainable fashion (this in fact has a larger impact on present-day quenched galaxies, see Paper II). In this sense, it would be interesting to investigate the outcome of both isolated and cosmological galaxy simulations and those with different feedback prescriptions, to see whether they can also produce such entangled fates between a galaxy's star forming activities and the CGM motion on larger scales. In particular, the bimodal kinematics of the CGM motion would be a key observation to test different scenarios.

\section*{Acknowledgements}
It has been a fully enjoyable research experience with a fantastic collaboration working on this paper series. We would also like to thank Drs. Cheng Li, Yong Shi, Yongquan Xue, Hong Guo, and an anonymous referee for their very constructive and insightful suggestions and comments. This work is partly supported by the National Key Research and Development Program of China (No. 2018YFA0404501 to SM), by the National Science Foundation of China (Grant No. 11821303, 11761131004 and 11761141012). DX also thanks the Tsinghua University Initiative Scientific Research Program ID 2019Z07L02017.

\section*{Data availability}
General properties of the galaxies in the IllustrisTNG Simulation is available from \url{http://www.tng-project.org/data/}. The rest of the data underlying the article will be shared on reasonable request to the corresponding authors.

\bibliographystyle{mnras}
\bibliography{ref}

\appendix
\section{Individual examples}
\label{apd:examples}

\begin{figure*}
\includegraphics[width=1\columnwidth]{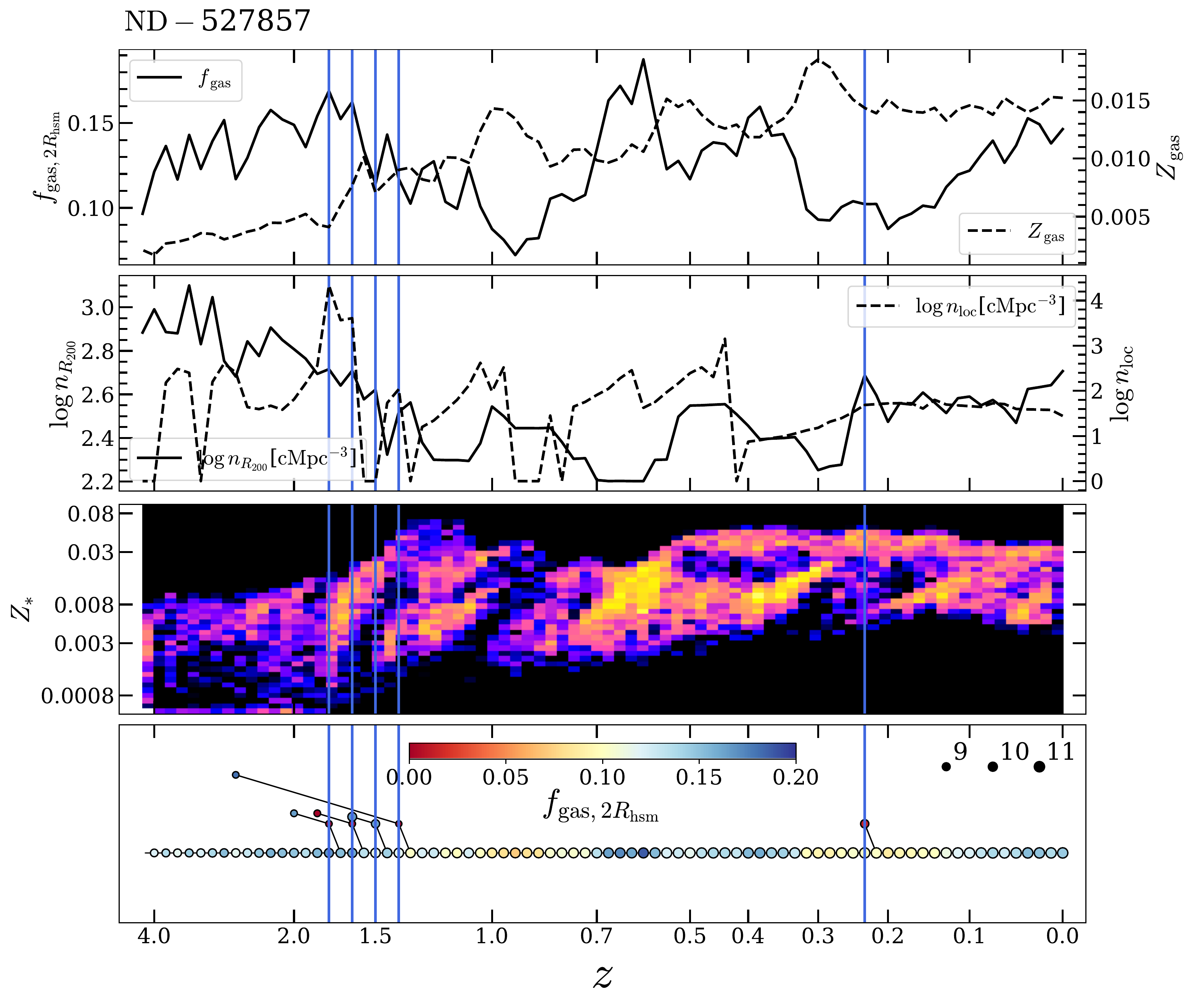}
\includegraphics[width=1\columnwidth]{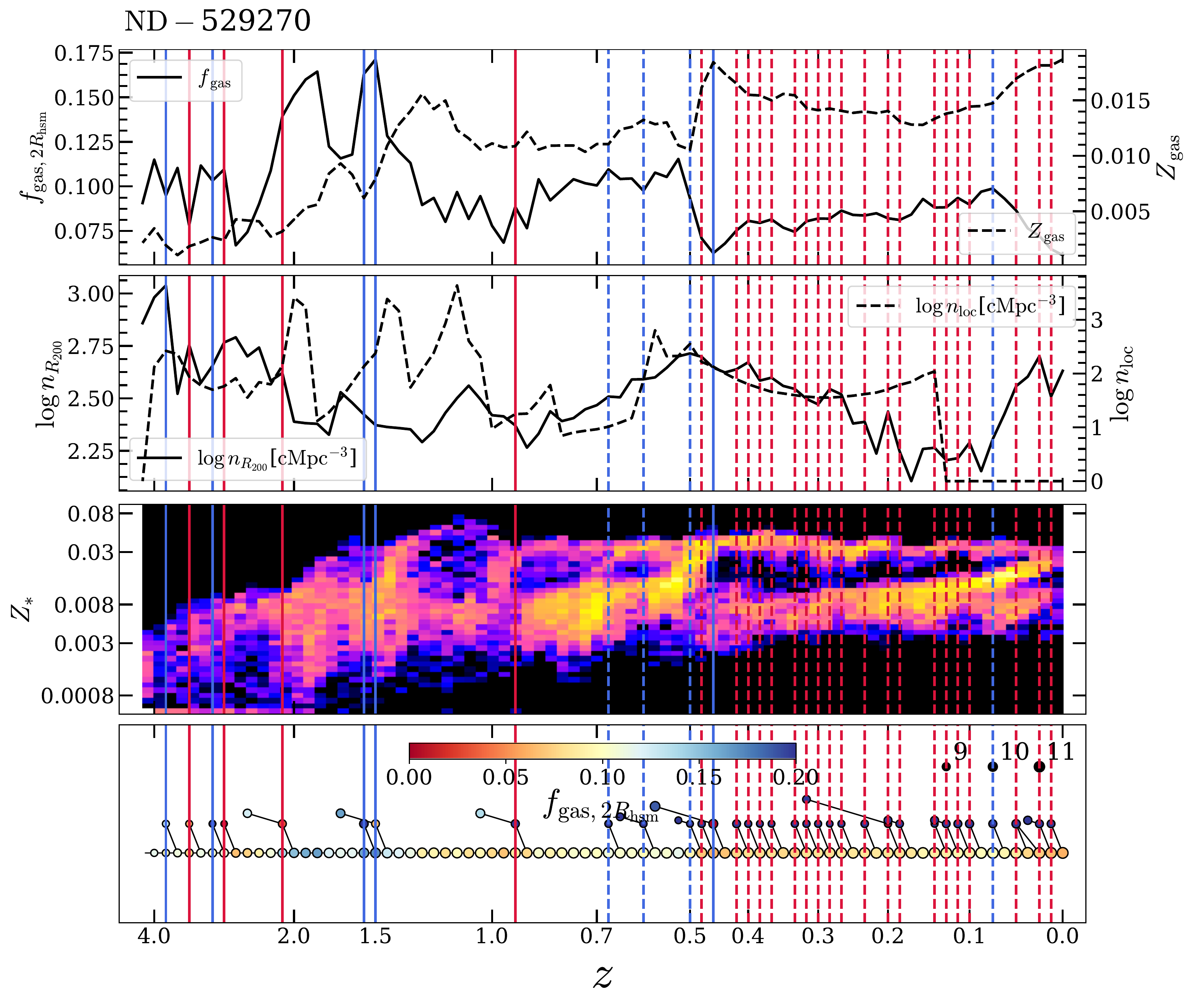}\\
\includegraphics[width=1\columnwidth]{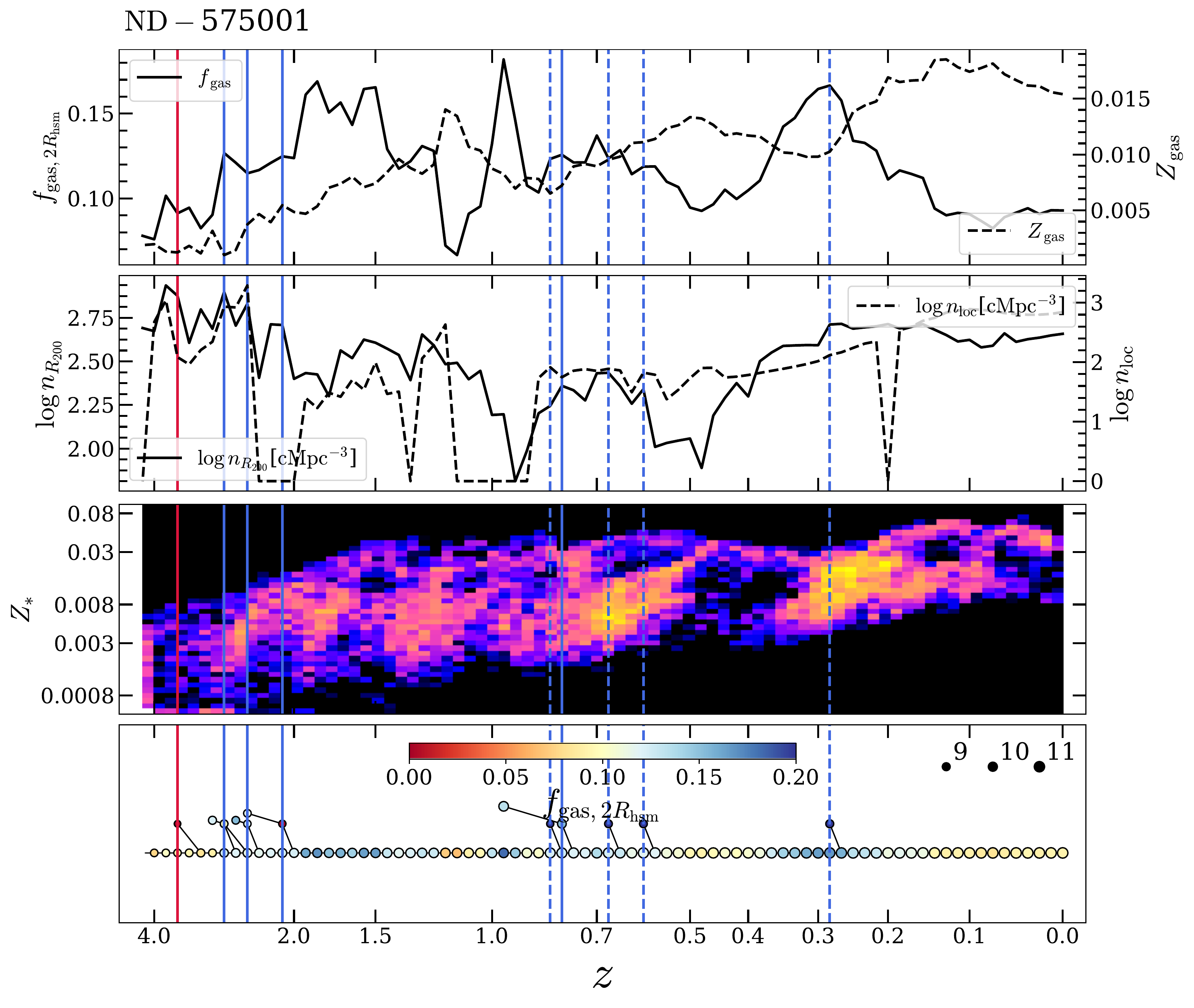}
\includegraphics[width=1\columnwidth]{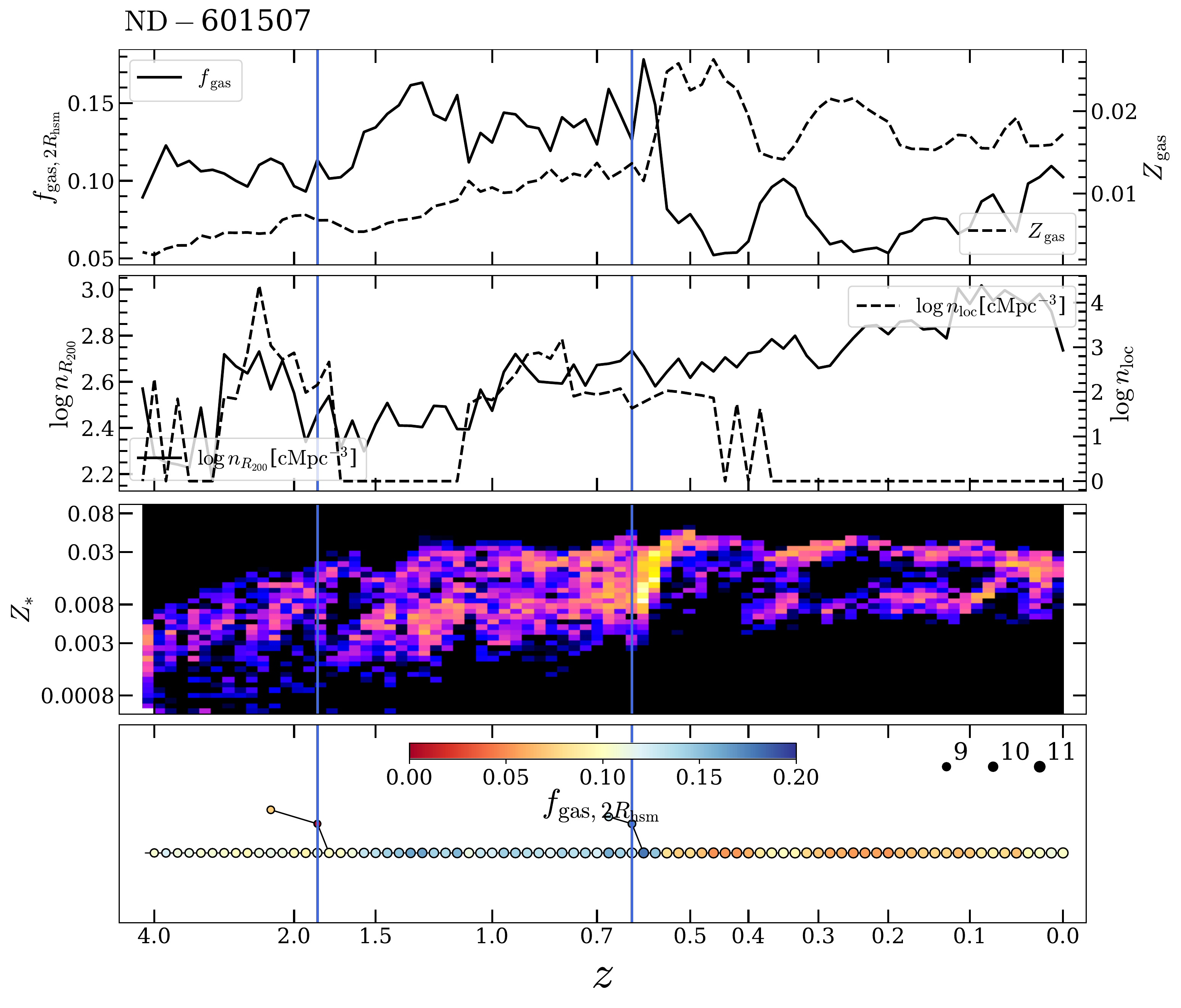}\\
\caption{From top to bottom, shown in each panel are the redshift evolution of {\bf (1)} the central gas fraction $f_{{\rm gas},2R_{\rm hsm}}$ (solid) and gas metallicity $Z_{{\rm gas},2R_{\rm hsm}}$ (dashed); {\bf (2)} galaxy neighbour counts $n_{R_{200}}$ (solid) and $n_{\rm loc}$ (dashed) of the environment (see text for detailed definitions); {\bf (3)}  star-forming age and metallicity phase-space intensity; and {\bf (4)} the galaxy's merger history (the same as in Fig.\,\ref{fig:Evolution_breathing}), for four present-day normal disc galaxies since z $\sim 4$. The galaxy IDs at $z=0$ are 527857, 529270, 575001 and 601507 from right to left and from top to bottom, respectively, which are displayed on the top left corners.}
\label{fig:example_track_ND}
\end{figure*}

\begin{figure*}
\includegraphics[width=1\columnwidth]{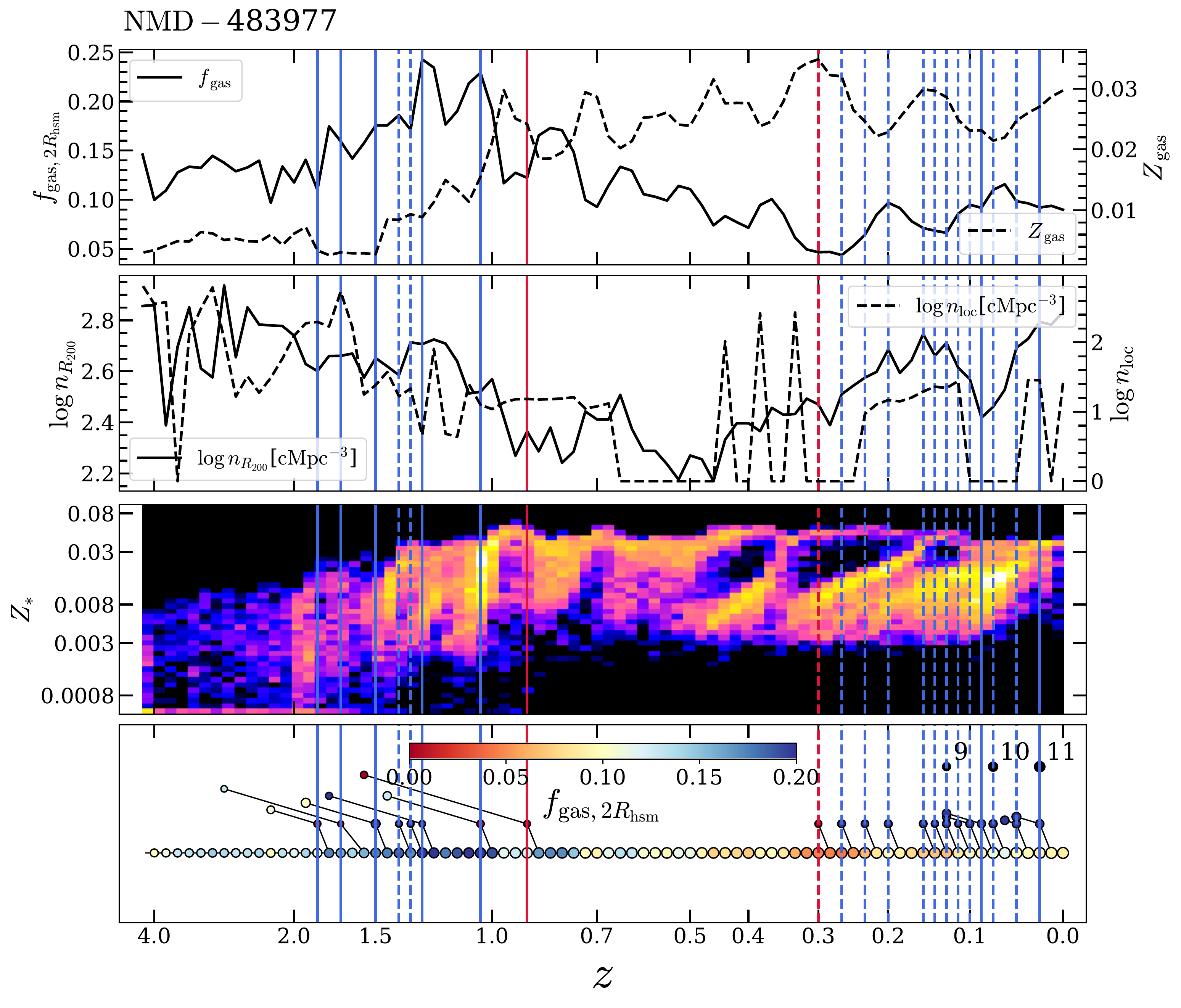}
\includegraphics[width=1\columnwidth]{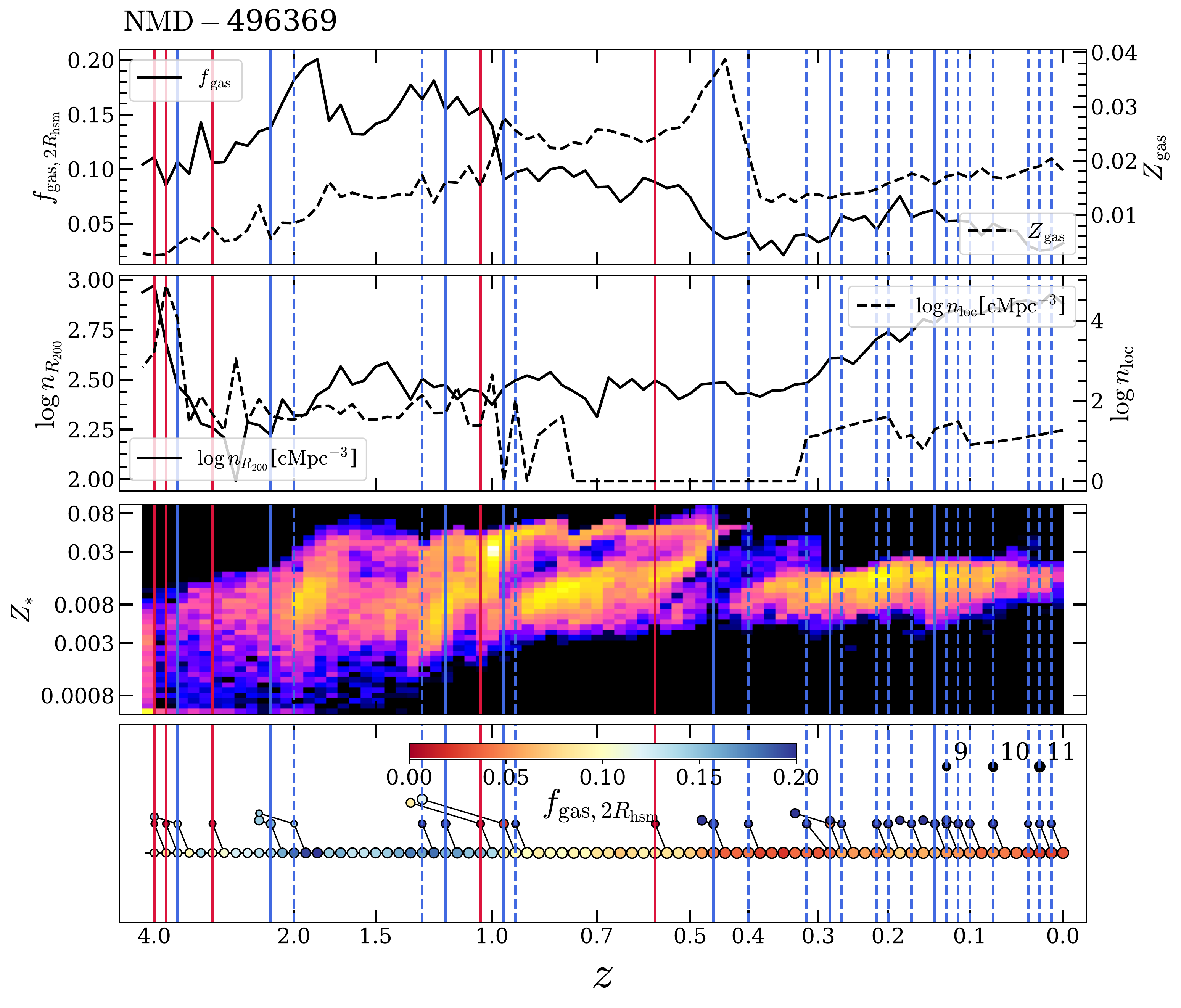}\\
\includegraphics[width=1\columnwidth]{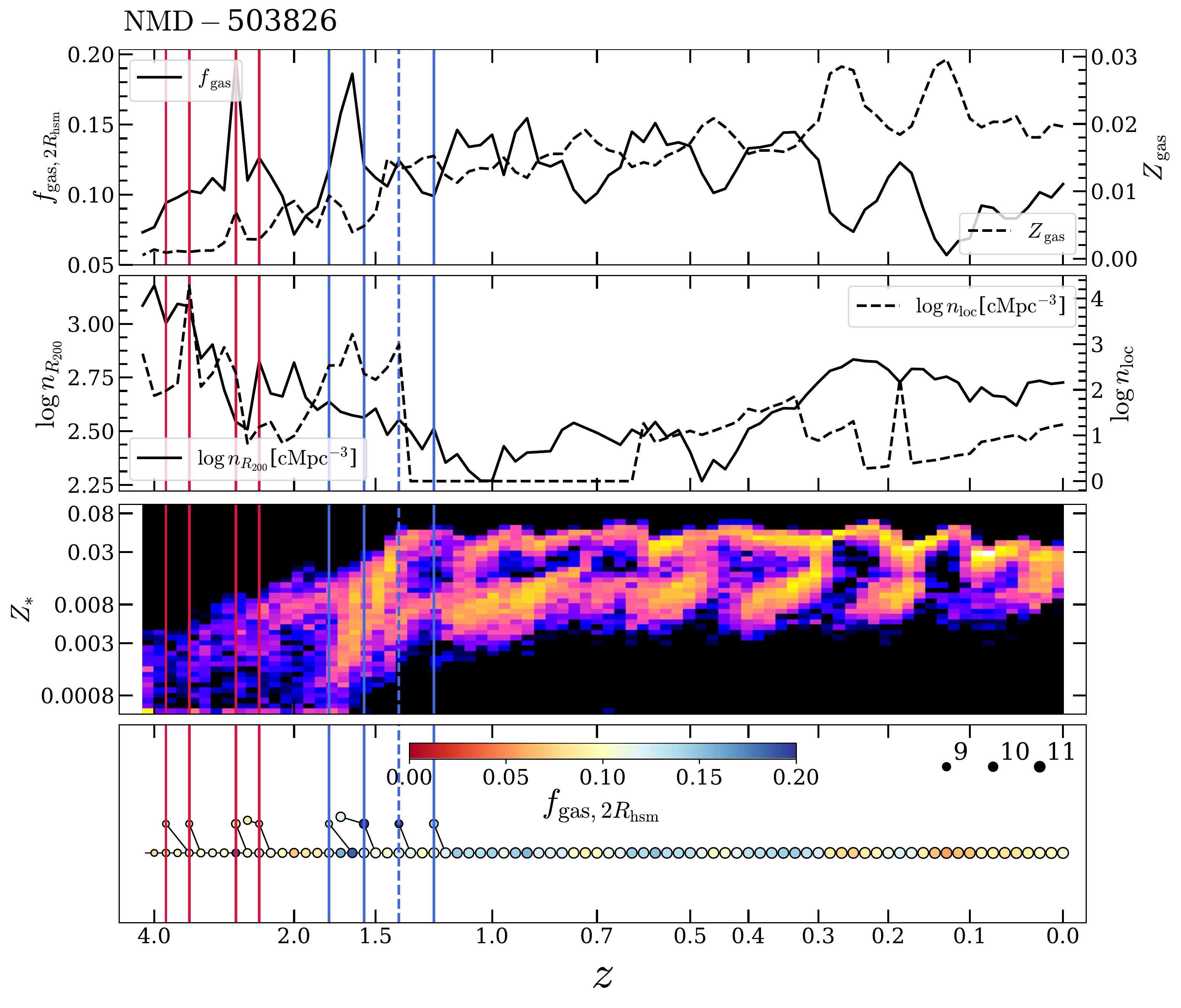}
\includegraphics[width=1\columnwidth]{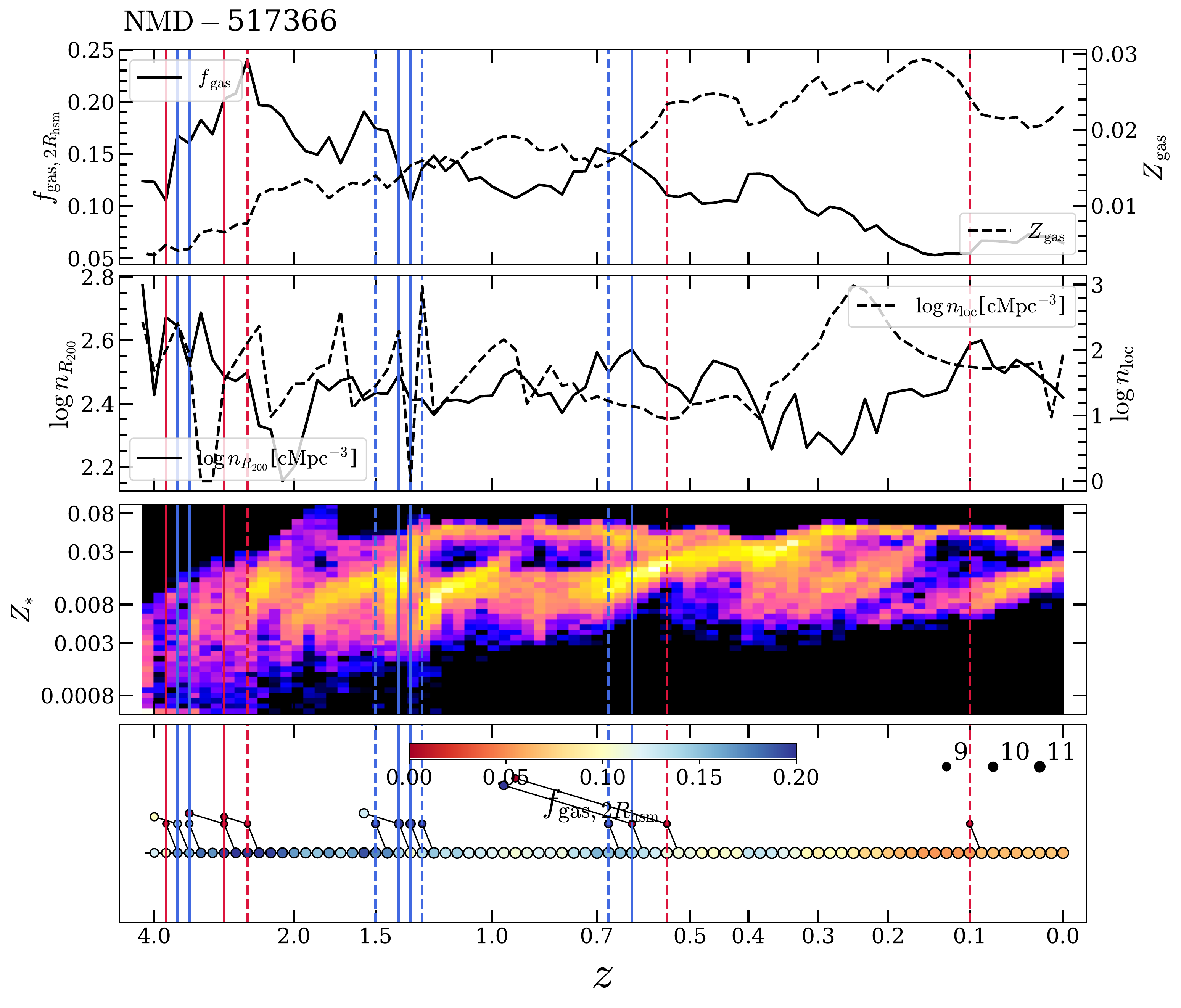}\\
\caption{Evolution of the same properties as in Fig.\,\ref{fig:example_track_ND}, for four present-day normal massive disc galaxies. The galaxy IDs at $z=0$ are 483997, 496369, 503826 and 517366 from right to left and from up to down, respectively.}
\label{fig:example_track_NMD}
\end{figure*}

In Figs.\,\ref{fig:example_track_ND} and \ref{fig:example_track_NMD}, from top to bottom, we present the redshift evolution of {\bf (1)} the central gas fraction $f_{{\rm gas},2R_{\rm hsm}}$ (solid) and gas metallicity $Z_{{\rm gas},2R_{\rm hsm}}$ (dashed); {\bf (2)} galaxy neighbour counts $n_{R_{200}}$ (solid) and $n_{\rm loc}$ (dashed) of the environment (see text for detailed definitions); {\bf (3)}  star-forming age and metallicity phase-space intensity; and {\bf (4)} the galaxy's merger history, of eight present-day star forming disc galaxies that have been traced back to $z=4$.

\label{lastpage}
\end{document}